\documentclass[12pt,preprint]{aastex}

\newcommand{\bdv}[1]{\mbox{\boldmath$#1$}}

\def\au{{\rm au}}

\def\masyr{{\rm mas}\,{\rm yr}^{-1}}
\def\kpc{{\rm kpc}}
\def\mas{{\rm mas}}

\def\muas{\mu{\rm as}}

\def\max{{\rm max}}

\def\rel{{\rm rel}}

\def\e{{\rm E}}
\def\bpi{{\bdv\pi}}

\def\bmu{{\bdv\mu}}

\def\bgamma{{\bdv\gamma}}

\def\btheta{{\bdv\theta}}

\begin{document}
\title{{Systematic KMTNet Planetary Anomaly Search, Paper II}:
{Six} New $q<2\times 10^{-4}$ Mass-ratio Planets}

\author{\textsc{
Kyu-Ha Hwang$^{1}$, 
Weicheng Zang$^{2}$,
Andrew Gould$^{3,4}$, 
Andrzej Udalski$^{5}$,
Ian A. Bond$^{6}$, 
Hongjing Yang$^{2}$,
Shude Mao$^{2,7}$ \\
(Lead Authors)\\
Michael D. Albrow$^{8}$, 
Sun-Ju Chung$^{1,9}$, 
Cheongho Han$^{10}$, 
Youn Kil Jung$^{1}$, 
Yoon-Hyun Ryu$^{1}$, 
In-Gu Shin$^{1}$, 
Yossi Shvartzvald$^{11}$, 
Jennifer C. Yee$^{12}$, 
Sang-Mok Cha$^{1,13}$, 
Dong-Jin Kim$^{1}$,
Hyoun-Woo Kim$^{1}$, 
Seung-Lee Kim$^{1,9}$, 
Chung-Uk Lee$^{1,9}$, 
Dong-Joo Lee$^{1}$,
Yongseok Lee$^{1,13}$, 
Byeong-Gon Park$^{1,9}$, 
Richard W. Pogge$^{4}$ \\
(KMTNet Collaboration)\\
Przemek Mr{\'o}z$^{14}$,
Radek Poleski$^{5,4}$,
Jan Skowron$^{5}$,
Micha{\l} K. Szyma{\'n}ski$^{5}$,
Igor Soszy{\'n}ski$^{5}$,  
Pawe{\l} Pietrukowicz$^{5}$,
Szymon Koz{\l}owski$^{5}$,
Krzysztof Ulaczyk$^{15}$,
Krzysztof A. Rybicki$^{5}$,
Patryk Iwanek$^{5}$,
Marcin Wrona$^{5}$,
Mariusz Gromadzki$^{5}$\\
(OGLE Collaboration)\\
Fumio Abe$^{16}$, 
Richard Barry$^{17}$, 
David P. Bennett$^{17,18}$, 
Aparna Bhattacharya$^{17,18}$, 
Hirosame Fujii$^{16}$
Akihiko Fukui$^{19,20}$, 
Yuki Hirao$^{21}$,
Yoshitaka Itow$^{16}$, 
Rintaro Kirikawa$^{21}$, 
Iona Kondo$^{21}$, 
Naoki Koshimoto$^{22,23}$, 
Brandon Munford$^{24}$, 
Yutaka Matsubara$^{16}$, 
Shota Miyazaki$^{21}$, 
Yasushi Muraki$^{16}$, 
Greg Olmschenk$^{17}$,
Cl{\'e}ment Ranc$^{25}$, 
Nicholas J. Rattenbury$^{24}$, 
Yuki K. Satoh$^{21}$,
Hikaru Shoji$^{21}$,
Stela Ishitani Silva$^{26,17}$,
Takahiro~Sumi$^{21}$,
Daisuke Suzuki$^{27}$, 
Paul J. Tristram$^{28}$, 
Atsunori Yonehara$^{29}$ \\
(The MOA Collaboration) \\
Xiangyu Zhang$^{2}$,
Wei Zhu$^{2}$,
Matthew T.\ Penny$^{30}$,
Pascal Fouqu\'e$^{31,32}$ \\
(The Tsinghua \& CFHT Microlensing Group) \\
} }

\affil{$^{1}$Korea Astronomy and Space Science Institute, Daejon
34055, Republic of Korea}

\affil{$^{2}$ Department of Astronomy and Tsinghua Centre for Astrophysics, 
Tsinghua University, Beijing 100084, China}

\affil{$^{3}$Max-Planck-Institute for Astronomy, K\"{o}nigstuhl 17,
69117 Heidelberg, Germany}

\affil{$^{4}$Department of Astronomy, Ohio State University, 140 W.
18th Ave., Columbus, OH 43210, USA}

\affil{$^{5}$Astronomical Observatory, University of Warsaw, 
Al.~Ujazdowskie~4, 00-478~Warszawa, Poland}

\affil{$^{6}$Institute of Natural and Mathematical Science, 
Massey University, Auckland 0745, New Zealand}

\affil{$^{7}$National Astronomical Observatories, Chinese Academy of Sciences, Beijing 100101, China}

\affil{$^{8}$University of Canterbury, Department of Physics and
Astronomy, Private Bag 4800, Christchurch 8020, New Zealand}

\affil{$^{9}$Korea University of Science and Technology, Korea, 
(UST), 217 Gajeong-ro, Yuseong-gu, Daejeon, 34113, Republic of Korea}

\affil{$^{10}$Department of Physics, Chungbuk National University,
Cheongju 28644, Republic of Korea}

\affil{$^{11}$Department of Particle Physics and Astrophysics, 
Weizmann Institute of Science, Rehovot 76100, Israel}

\affil{$^{12}$ Center for Astrophysics $|$ Harvard \& Smithsonian, 60 Garden
St., Cambridge, MA 02138, USA}

\affil{$^{13}$School of Space Research, Kyung Hee University,
Yongin, Kyeonggi 17104, Republic of Korea}

\affil{$^{14}$Division of Physics, Mathematics, and Astronomy, 
California Institute of Technology, Pasadena, CA 91125, USA}

\affil{$^{15}$Department of Physics, University of Warwick, 
Gibbet Hill Road, Coventry, CV4~7AL,~UK}

\affil{$^{16}$Institute for Space-Earth Environmental Research, 
Nagoya University, Nagoya 464-8601, Japan}

\affil{$^{17}$Code 667, NASA Goddard Space Flight Center, 
Greenbelt, MD 20771, USA}

\affil{$^{18}$Department of Astronomy, University of Maryland, 
College Park, MD 20742, USA}

\affil{$^{19}$Department of Earth and Planetary Science, Graduate School of Science, 
The University of Tokyo, 7-3-1 Hongo, Bunkyo-ku, Tokyo 113-0033, Japan}

\affil{$^{20}$Instituto de Astrof{\'i}sica de Canarias, 
V{\'i}a L{\'a}ctea s/n, E-38205 La Laguna, Tenerife, Spain}

\affil{$^{21}$Department of Earth and Space Science, Graduate School of Science, 
Osaka University, Toyonaka, Osaka 560-0043, Japan}

\affil{$^{22}$Department of Astronomy, Graduate School of Science, 
The University of Tokyo, 7-3-1 Hongo, Bunkyo-ku, Tokyo 113-0033, Japan}

\affil{$^{23}$National Astronomical Observatory of Japan, 2-21-1 Osawa, 
Mitaka, Tokyo 181-8588, Japan}

\affil{$^{24}$Department of Physics, University of Auckland, 
Private Bag 92019, Auckland, New Zealand}


\affil{$^{25}$Sorbonne Universit\'e, CNRS, UMR 7095, Institut d'Astrophysique de Paris, 98 bis bd Arago, 75014 Paris, France}

\affil{$^{26}$Department of Physics, The Catholic University of America, Washington, DC 20064, USA}

\affil{$^{27}$Institute of Space and Astronautical Science, 
Japan Aerospace Exploration Agency, Kanagawa 252-5210, Japan}

\affil{$^{28}$University of Canterbury Mt. John Observatory, 
P.O. Box 56, Lake Tekapo 8770, New Zealand}

\affil{$^{29}$Department of Physics, Faculty of Science, 
Kyoto Sangyo University, Kyoto 603-8555, Japan}

\affil{$^{30}$Department of Physics and Astronomy, Louisiana State University, Baton Rouge, LA 70803 USA}

\affil{$^{31}$CFHT Corporation, 65-1238 Mamalahoa Hwy, Kamuela, Hawaii 96743, USA}

\affil{$^{32}$Universit\'e de Toulouse, UPS-OMP, IRAP, Toulouse, France}

\begin{abstract}

We apply the automated AnomalyFinder algorithm of 
Paper I \citep{ob191053} to 2018-2019 light curves from 
the $\simeq 13\,{\rm deg}^2$ covered by the six KMTNet prime fields,
with cadences $\Gamma \geq 2\,{\rm hr}^{-1}$.  We find a total of
{11} planets with mass ratios $q<2\times 10^{-4}$, including {six} 
newly discovered 
planets, one planet that was reported in Paper I, and recovery of four
previously discovered planets.  One of the new planets, OGLE-2018-BLG-0977Lb,
is in a planetary-caustic event, while the other {five}
(OGLE-2018-BLG-0506Lb, OGLE-2018-BLG-0516Lb, OGLE-2019-BLG-1492Lb, 
 KMT-2019-BLG-0253, {and KMT-2019-BLG-0953}) are revealed by a ``dip''
in the light curve as the source crosses the host-planet axis on the
opposite side of the planet.  These subtle signals were missed in
previous by-eye searches.  The planet-host separations (scaled to the
Einstein radius), $s$, and planet-host mass ratios, $q$, are, respectively,
$(s,q\times 10^5) = (0.88, 4.1)$, 
$(0.96\pm 0.10, 8.3)$, 
$(0.94\pm 0.07, 13)$, 
$(0.97\pm 0.07, 18)$, 
$(0.97\pm0.04,4.1)$, 
and
$(0.74,18)$, 
where the ``$\pm$'' indicates a discrete degeneracy.
The {11} planets are spread out over the range $-5<\log q < -3.7$.
Together with the two planets previously reported with $q\sim 10^{-5}$
from the 2018-2019 non-prime KMT fields, this result suggests
that planets toward the bottom of this mass-ratio range may be more common than
previously believed.

\end{abstract}


{\section{{Introduction}
\label{sec:intro}}

By the end of 2005, four microlensing planets had been discovered,
of which two were in the low planet-host mass-ratio ($q<10^{-4}$) regime
\citep{ob05390,ob05169}.  Twelve years later, there were only seven
such discoveries \citep{ob171434} out of a total of more than 60.
Moreover, none of these had mass ratios below 
$q_{\rm thresh,2018.0}=4.3\times 10^{-5}$.
The absence of low-$q$ planets in a more restricted, but homogeneously
selected sample, led \citet{suzuki16} to argue for a break in the 
mass-ratio function at about $q_{\rm break}=17\times 10^{-5}$, albeit
with a large error bar.  That is, they confirmed the earlier finding
by \citet{ob07368} that above the break,
the mass-ratio function was a falling power law toward higher mass ratios.
However, below the break, they found that this function was either flat, or more
likely falling sharply toward lower mass ratios.  Based on a larger,
though inhomogeneous sample, \citet{kb170165} found the break to be
a factor three lower, $q_{\rm break}=5.6\times 10^{-5}$, but still
consistent with \citet{suzuki16} within the latter's error bar.

Since the \citet{ob171434} study (and prior to this paper), seven
additional $q\leq 10^{-4}$ planets have been discovered.  Five of these
seven have mass ratios below the pre-2018 threshold, 
$q_{\rm thresh,2018.0}=4.3\times 10^{-5}$, and these five were discovered in 
2018--2020 data.  Although these 14 planets 
were not homogeneously selected, it is unlikely that this
sample is strongly affected by publication bias.  That is, $q<10^{-4}$
planets are relatively rare and are very likely to be published soon after
discovery.  Only one of the seven recent low-$q$ planets was submitted
for publication substantially after discovery, OGLE-2015-BLG-1670
\citep{ob151670}.  

The very different discovery patterns before and after 2018.0 strongly
suggest a systematic difference in the discovery process
beginning in 2018.  See Table~\ref{tab:sel}.
One possibility is the upgrade in the
Korea Microlensing Telescope Network (KMTNet, \citealt{kmtnet}) online
photometry in that year.  KMT data played a major or decisive role
in all five 2018--2020 discoveries, and many of these planets
would have been missed with substantially inferior photometry.
For example, KMT-2018-BLG-0029 \citep{kb180029}
had been the focus of considerable
interest during the season because it was targeted for {\it Spitzer}
observations, yet the few-hour planetary signal at the peak was
not noticed until the end-of-year re-reductions, when the new pipeline
results were manually reviewed.

Regardless of the exact cause, the changing pattern of discoveries
before and after 2018.0 calls for re-evaluation of the conclusions
regarding the behavior of the mass-ratio function in the
$\log q\la -4$ regime.  Ultimately, this
can only be done on the basis of homogeneous selection of planets,
which was a key characteristic of the original \citet{suzuki16} study.

 \begin{deluxetable}{lccr}
 \tablecolumns{4} \tablewidth{0pc}
 \tablecaption{\textsc{Summary of Microlensing Planets $q<10^{-4}$}}
 \tablehead{\colhead{Years} & 
\colhead{$q<4.3\times 10^{-5}$} &
\colhead{$q>4.3\times 10^{-5}$} &
 \colhead{Total} }
 \startdata
2003-2017 & 0 & 8 & 8 \\
2018-2020 & 5 & 1 & 6 \\
\hline
All       & 5 & 9 & 14 \\
 \enddata
 \label{tab:sel}
 \end{deluxetable}

\citet{ob190960} outlined one such approach: make tender-loving care (TLC)
reductions of all KMT events meeting a magnification threshold, e.g.,
$A_\max >20$, and subject these to automated planet searches.  This
would require improved efficiency of the TLC pipeline to re-reduce so
many events.

\citet{kb200414} outlined a second approach: intensive follow-up observations
of high-magnification events in low-cadence $\Gamma \leq 0.4\,{\rm hr}^{-1}$
KMT fields.  This would be similar to the approach formerly carried out by
the Microlensing Follow Up Network ($\mu$FUN, \citealt{gould10}), but
with much more efficient target selection due to continuous KMT coverage.

\citet{ob191053} (Paper I) initiated a third approach: subject
the residuals from single-lens single-source (1L1S) fits to
the end-of-year-pipeline light curves to a search for ``bumps''
or ``dips'' based on a modified version of the KMT EventFinder
algorithm \citep{eventfinder,gould96}.  At that time, the \citet{ob191053} 
approach could
only be applied to 2019 data because the 2016--2017 online reductions
were not of sufficiently high quality and the 2018 photometry files
lacked some auxiliary data needed for this method.  However, the 2018
files have already been updated and work on the 2016-2017 files is
in progress.  Note that all three
methods are complementary and all could be used simultaneously.

As reported by \citet{ob191053}, their method
recovered the planets in 
KMT-2019-BLG-0842 \citep{kb190842} and
OGLE-2019-BLG-0960 \citep{ob190960}, 
i.e., the two previous low-$q$ planets from 2019.  And they also reported
the discovery of the lowest mass-ratio planetary-caustic\footnote{
There were three planets with comparable mass ratios
KMT-2018-BLG-0029 \citep{kb180029},
OGLE-2019-BLG-0960 \citep{ob190960}, and 
KMT-2020-BLG-0414 \citep{kb200414}.  However, all three were detected
via the resonant channel, which, together with the ``near-resonant'' channel,
account for about 3/4 or all published microlensing planets.} 
event to date:
OGLE-2019-BLG-1053.  Hence, there are good prospects for establishing
a large homogeneous sample of $q<10^{-4}$ planets by applying this technique
to several years of KMT data.  Here, we begin a program of detailed
analysis and publication of all $q<2\times 10^{-4}$ planets found from 
application of the \citet{ob191053} method.  This will 
enable a systematic study of the planet mass-ratio function
in the regime of the hypothesized break.

We briefly outline the steps leading up to these publications and the
steps that are expected to follow.  As discussed by \citet{ob191053}, 
their automated AnomalyFinder selects potential anomalies based on
purely objective criteria applied to KMT light curves.  The criteria
used here are identical except that we extend the effective-timescale 
range down to $t_{\rm eff} \geq 0.05\,$day.
Three operators (W.Zang, H. Yang and W. Zhu) then manually 
and independently select genuine anomalies (or anomalies that could be genuine)
based on machine-generated displays.  If there are counterparts to the KMT
event on the OGLE and/or MOA web pages, then these light curves
are checked to determine whether they contradict the apparent KMT
anomaly.  If so, the anomaly is rejected.  If there are no
OGLE or MOA data that overlap the anomaly, its reality 
is checked by visual inspection of the images.  {Ultimately, we
expect to publish all planets (defined as $q<0.03$) derived from this
search, which will eventually be extended to all KMTNet fields and
at least the 2016--2019 seasons.}

The present paper is based on a search of the KMT database that is restricted
to 2018--2019 and to the three pairs of overlapping KMT fields 
BLG01/BLG41, BLG02/BLG42, and BLG03/BLG43.  Each of these fields has a nominal
cadence $\Gamma=2\,{\rm hr}^{-1}$, so the combined cadence is usually
$\Gamma=4\,{\rm hr}^{-1}$.  The first restriction is determined by the
current state of the photometry files (see above).  The second is based
on maintaining a stable work flow in what will be a very big project.

In particular, each event that is selected for publication must be subjected
to detailed analysis, so their publication in groups does not substantially
reduce either the amount of work required or the length of the descriptions.

The long-term goal is to measure the mass-ratio function for the
full planetary range $q< 0.03$.
To do so, we must measure the sensitivity of the selection process
by injecting planets into KMT light curves.  Then two questions will be
asked.  First, is the anomaly selected by the machine criteria?
Second, would this machine-selected candidate ultimately be published?
The answer to the first question is unambiguous.  The second question
is more difficult because it involves several steps that cannot be
mimicked for each simulated event, such as re-reduction of the data and 
systematic searches for alternate models.  This analysis is currently
in progress (Y.K.~Jung et al., in prep).

Here, we present {six} planets, all of which ultimately satisfy 
$q<2\times 10^{-4}$.
{To construct the present sample, we
manually review all anomalous light curves and tentatively classify
them by eye into one of five categories 
``planet'', ``planet/binary'', ``binary/planet'', ``binary'', and
``finite source''.  
Among the ``planets'', we do preliminary 2L1S modeling of all events that
could plausibly be $q<10^{-3}$, using end-of-season pipeline data.
Here, $n$L$m$S means $n$ lenses and $m$ sources.  
One of these, OGLE-2018-BLG-0383 (with $q=2.1\times 10^{-4}$), 
was separated out at an early stage as part of an investigation
of wide-orbit planets \citep{ob180383}.  
For those with $q<3\times 10^{-4}$ among the remainder, } 
the KMT data are re-reduced with TLC and then fully analyzed.  

{
All of the other ``planets'' as well as all of the events in the
other four categories are then cross-matched against events that are
published or are known to us to be in preparation.  We then cross match
against the very large personal modeling archive of co-author C.\ Han.
Combined, these cross matches cover about 75\% of all the 
AnomalyFinder detections.  They give us not only good estimates of $q$
for these events, but also permit
us to check the robustness of our by-eye classifications.  Based on this
investigation, it is likely that all of the $q<2\times 10^{-4}$ planets
in the AnomalyFinder sample are contained in the present work.  However,
we cannot be certain until we complete our systematic investigation of
all the anomalous events, which is currently underway.
}

{\section{{Observations}
\label{sec:obs}}

As outlined in Section~\ref{sec:intro},
all {six} planets described in this paper were identified in searches of
KMT events from its ``prime fields'', 
with nominal cadences $\Gamma\geq 2\,{\rm hr}^{-1}$, either by the 
AlertFinder \citep{alertfinder} or the post-season EventFinder
\citep{eventfinder}.  KMTNet observes
from three identical 1.6m telescopes, each equipped with
a $(2^\circ\times 2^\circ)$ camera at CTIO in Chile (KMTC),
SAAO in South Africa (KMTS), and SSO in Australia (KMTA).
KMTNet observes primarily in the $I$ band, but
in 2018 and 2019, every tenth such observation was complemented by
one in the $V$ band.

{Five of the six} events were independently discovered by 
the Optical Gravitational
Lensing Experiment (OGLE, \citealt{ews1,ews2}), using its 1.3m
telescope with $1.4\,{\rm deg}^2$ camera at Las Campanas Observatory
in Chile.  OGLE also observed primarily in the $I$ band, with some
$V$ exposures as well.

Two of the {six} events were independently discovered by the Microlensing
Observations in Astrophysics (MOA, \citealt{ob03235}) collaboration,
using their 1.8m telescope with $2.2\,{\rm deg}^2$ camera at Mt.\ John
in New Zealand, using their
$R_{\rm MOA}$ filter, which is roughly the sum of the Cousins $R$ and $I$ bands.

Table~\ref{tab:names} gives
the catalog names in order of discovery for each event
in order to permit information to be easily traced.  However, after that,
we generally use the first-discovery name.  Table~\ref{tab:names} also
presents the observational cadences $\Gamma$, as well as the discovery
dates and the sky locations.

In addition, {OGLE observed the location of KMT-2019-BLG-0953 and}
MOA observed the location of the OGLE-2019-BLG-1492
field, although {they did not issue alerts for these events}. 
We therefore also incorporate these {OGLE and} MOA data into the analysis.

Although all {six} events were alerted in real time, 
to the best of our knowledge, there were follow-up observations for only one:
OGLE-2018-BLG-0977.  These were from the 3.6m Canada-France-Hawaii Telescope
(CFHT) and were at a cadence of about one per night.

The data were reduced using variants of 
difference image analysis (DIA,\citealt{tomaney96,alard98}), as realized by
\citet{albrow09} (KMT), \citet{wozniak2000} (OGLE), \citet{bond01} (MOA),
and \citet{cfht-phot} (CFHT).

{\section{{Light-curve Analyses}
\label{sec:anal}}

{\subsection{{Preamble}
\label{sec:anal-preamble}}

There are several features of the light-curve analyses that are common
to all {six} events, which we present in this preamble.  

All {six} events exhibit short perturbations on otherwise standard
\citet{pac86} 1L1S light curves, which are characterized by three
parameters (in addition to two flux parameters for each observatory).
These are $(t_0,u_0,t_\e)$, i.e., the time of lens-source closest approach,
the impact parameter (in units of the Einstein radius, $\theta_\e$),
and the Einstein timescale.

In each case, the anomaly can be localized fairly precisely by eye at
$t_{\rm anom}$, which then yields the offset from the peak, $\tau_{\rm anom}$
and the offset from the host, $u_{\rm anom}$, both in units of $\theta_\e$,
as well as the angle $\alpha$ of the source-lens motion relative to
the binary axis \citep{gouldloeb},
\begin{equation}
\tau_{\rm anom} = {\Delta t_{\rm anom}\over t_\e}
\equiv {t_{\rm anom} - t_0\over t_\e};
\qquad 
u_{\rm anom}^2 = \tau_{\rm anom}^2 + u_0^2;
\qquad
\tan\alpha = -{u_0\over\tau_{\rm anom}}.
\label{eqn:tau_anom}
\end{equation}
Note that this also implies that $u_{\rm anom} = |u_0/\sin\alpha|$.

We define the quantities $s^\dagger_\pm$ by
\begin{equation}
s^\dagger_\pm = {\sqrt{u_{\rm anom}^2 + 4} \pm u_{\rm anom}\over 2}.
\label{eqn:sdagger}
\end{equation}
Note that $s^\dagger_- = 1/s^\dagger_+$ and 
$u_{\rm anom} = s^\dagger_+ - s^\dagger_-$.
If the source crosses the minor-image (triangular) planetary caustics, 
then we expect $s\simeq s^\dagger_-$, and if it crosses the major 
image (quadrilateral) planetary caustic, 
then we expect $s\simeq s^\dagger_+$.  Here, $s$ is the
projected planet-host separation scaled to $\theta_\e$.  However, only
one of the {six} events analyzed here (OGLE-2018-BLG-0977) has such
caustic-crossing features.  For the rest, we may expect that the
event {may be} subject to the inner/outer degeneracy identified by 
\citet{gaudi97}.  That is, for minor-image perturbations, which
are generally characterized by a ``dip'', roughly the same anomaly 
can in principle be produced by the source passing ``inside'' the
planetary caustics (closer to the central caustic) or ``outside''.
Then we may find two solutions, which obey,
\begin{equation}
s^\dagger_- = {s_{\rm inner} + s_{\rm outer}\over 2}
\qquad {\rm (minor\ image)}.
\label{eqn:s_average}
\end{equation}
For major-image perturbations, which are generally characterized by a ``bump'',
a similar formula applies for $s^\dagger_+$.  However, in the present
paper, all {six} planets give rise to minor-image perturbations (``dips'').

\citet{gouldloeb} also show how to make by-eye estimates of $q$ for
events in which the source crosses a planetary caustic.  Here, we
present an extension of their method to estimate $q$ for the
case of non-caustic-crossing minor-image
perturbations caused by low mass-ratio planets.

We first focus on caustic-crossing events.  In this case, there
is a de-magnification trough between the two flanking caustics, with
separation (using notation from \citealt{han06}),
\begin{equation}
\Delta u \simeq 2\eta_{c,1} - \Delta\eta_c 
= 4\sqrt{q(s^{-1} - s)\over s} = 4\sqrt{q {u_{\rm anom}\over s}}.
\label{eqn:han1}
\end{equation}
Therefore, the duration of the dip in the light curve between the two
caustics is $\Delta t_{\rm dip} = (\Delta u/\sin\alpha)t_\e$, implying
\begin{equation}
q = \biggl({\Delta t_{\rm dip}\over 4\, t_\e}\biggr)^2
{s\sin^2\alpha\over u_{\rm anom}} 
= \biggl({\Delta t_{\rm dip}\over 4\, t_\e}\biggr)^2
{s\over |u_0|}|\sin^3\alpha| .
\label{eqn:qeval}
\end{equation}
For the case of source trajectories that miss the caustic,
but still experience a dip (by which the planet is detected),
we apply the same formula, but with $s\rightarrow s^\dagger_-$,
i.e., implicitly assuming that the width of the dip is approximately
the same as the separation between the caustics. 
As shown by Figure~1 of \citet{chung11}, this approximation
holds best for low mass-ratio lenses $q\la 2\times 10^{-4}$, which is the
main focus of interest here.
We do not expect this mass-ratio estimate
to be extremely accurate simply
because the dip does not itself have precisely defined edges (except
in the region of the trough that lies between the two triangular caustics).
Moreover, $\Delta t_{\rm dip}$ enters quadratically into 
Equation~(\ref{eqn:qeval}).
Thus, by contrast with Equations~(\ref{eqn:tau_anom}) and (\ref{eqn:sdagger}),
which should be quite precise, we expect that Equation~(\ref{eqn:qeval})
should be accurate at the factor $\sim 2$ level.

Nevertheless, Equation~(\ref{eqn:qeval}) can be quite powerful.  Note
in particular that for events in which the anomaly occurs at magnifications
of at least a few, $A_{\rm anom}\ga 5$ and for which the blending is small
compared to the magnified flux, we have 
$\sin\alpha = u_0/u_{\rm anom} \simeq A_{\rm anom}/A_\max \simeq 
10^{-0.4\Delta I}$, where $\Delta I$ is the magnitude offset between
the peak and the anomaly.  For such cases 
$s^\dagger_- \rightarrow 1 - 0.5/A_{\rm anom} \simeq 1$, and so
Equation~(\ref{eqn:qeval}) can be approximated by
\begin{equation}
q\rightarrow {10^{-1.2\Delta I}\over u_0}
\biggl({\Delta t_{\rm dip}\over 4\, t_\e}\biggr)^2;
\qquad {\rm (limiting\ form)}.
\label{eqn:rapidq}
\end{equation}

We generally search for seven-parameter ``static'' (i.e., without lens orbital
motion or microlens parallax effects) 2L1S solutions on an $(s,q)$ grid,
characterized by $(t_0,u_0,t_\e,s,q,\alpha,\rho)$, where $\rho$ is the 
angular source radius scaled to $\theta_\e$, i.e., $\rho=\theta_*/\theta_\e$.
In the grid search, $(s,q)$ are held fixed at the grid values
while $(t_0,u_0,t_\e,\rho)$ are allowed to vary in a Markov chain
Monte Carlo (MCMC).
{We use the advanced contour integration code \texttt{VBBinaryLensing}
\citep{bozza10,bozza18} to calculate the magnification of the
2L1S model, and we identify the best-fit solution via the Markov chain
Monte Carlo (MCMC) method \texttt{emcee} \citep{emcee}.  }
The three Paczy\'nski parameters 
$(t_0,u_0,t_\e)$ are seeded at their 1L1S values, and $\rho$ is seeded
using the method of \citet{gaudi02}
at a value near $\rho=\theta_{*,\rm est}/\mu_{\rm typ}t_\e$, where
$\mu_{\rm typ}\equiv 6\,\masyr$ is a typical value of the lens-source
relative proper motions, $\mu_\rel$, and $\theta_{*,\rm est}$ is estimated
based on the value of $I_s$ from the 1L1S fit and the KMT-website extinction.
For example, $\theta_{*,\rm est}=0.6\,\muas$ for $I_s-A_I = 18.65$ (Sun-like 
star) and
$\theta_{*,\rm est}=6\,\muas$ for $I_s-A_I = 14.5$ (clump-like star).  
The final parameter, $\alpha$
is seeded at a grid of values around the unit circle, and it is either held
fixed or allowed to vary with the chain, depending on circumstances.

After finding one or more local minima on the $(s,q)$ plane, each local
is further refined by allowing all seven parameters to vary in an MCMC.

If it is suspected that microlens parallax effects can be detected or 
constrained, then four additional parameters are initially added, 
$\bpi_\e=(\pi_{\e,N},\pi_{\e,E})$ and $\bgamma=((ds/dt)/s,d\alpha/dt)$.
Here, $\bpi_\e$ parameterizes the effects of Earth's orbital motion
on the light curve \citep{gould92,gould00,gould04},
\begin{equation}
\bpi_\e = {\pi_\rel\over\theta_\e}\,{\bmu_\rel\over\mu_\rel};
\qquad 
\theta_\e = \sqrt{\kappa M\pi_\rel};
\qquad
\kappa\equiv {4 G\over c^2\au}\simeq 8.14{\mas\over M_\odot},
\label{eqn:pie}
\end{equation}
where $M$ is the lens mass and $(\pi_\rel,\bmu_\rel)$ are the lens-source
relative (parallax, proper motion), while $(ds/dt,d\alpha/dt)$ are the
first derivatives on lens orbital motion.  The four
parameters must be introduced together,
at least initially, because the orbital motion of the lens can mimic
the effects on the light-curve of orbital motion of 
Earth \citep{mb09387,ob09020}.  Lens orbital motion can be poorly constrained
by the light curve, so that we generally impose a constraint $\beta<0.8$
where $\beta$ is the absolute value of the ratio of transverse
kinetic to potential energy \citep{eb2k5,ob05071b},
\begin{equation}
\beta = \bigg|{\rm KE\over PE}\bigg|_\perp = 
{\kappa M_\odot {\rm yr}^2\over 8\pi^2}{\pi_\e\over\theta_\e}\gamma^2
\biggl({s\over \pi_\e + \pi_S/\theta_\e}\biggr)^3 ,
\label{eqn:beta}
\end{equation}
and where $\pi_S$ is the source parallax.  Strictly speaking $\beta$ is
only physically constrained to be $\beta<1$, but combinations of
orbits and viewing
angles that would generate $\beta>0.8$ are extremely rare.

When (as usual) the parallax vector $\bpi_\e$ is evaluated in the
geocentric frame, it often has elongated, nearly straight contours
whose minor axis is aligned with the projected position of the Sun at $t_0$,
i.e., the (negative of the) direction of the Sun's instantaneous
apparent direction of acceleration on the plane of the sky.
This is because $\pi_{\e,\parallel}$, the component of
$\bpi_\e$ that is aligned to this acceleration,
induces an asymmetry in the light curve, which leads to strong constraints,
whereas $\pi_{\e,\perp}$ induces more subtle, symmetric distortions
\citep{gmb94,smp03,gould04}.  Because the right-handed coordinate
system $(\pi_{\e,\parallel},\pi_{\e,\perp})$
is rotated with respect to the (North, East) equatorial coordinates,
the tightness of the error ellipse is often not fully reflected when
$(\pi_{\e,N},\pi_{\e,E})$ and their errors are tabulated in papers.
In this paper, we therefore also tabulate $(\pi_{\e,\parallel},\pi_{\e,\perp})$
and their errors in addition to those of $(\pi_{\e,N},\pi_{\e,E})$.
For this purpose, we define $\psi$ as the angle of the minor axis
of the parallax ellipse (North through East) rather than the angle of
the projected position of the Sun.  Of course, there are two such
directions, separated by $180^\circ$.  We choose the one closer
to the Sun's apparent acceleration on the sky.  In all four events 
for which the parallax is measurable in the current work, $\psi$ is the same 
as the Sun's acceleration within one or two degrees.  See, e.g., Figure~3
of \citet{mb03037} for an explicit example of the coordinate system.

Even when these so-called ``one-dimensional (1-D) parallax'' measurements
fail to constrain the magnitude $|\bpi_\e|=\pi_\e$, which is what directly enters
the mass and distance estimates, they can interact with the 
Bayesian priors from a Galactic model to constrain the final
estimates of these quantities substantially better than $\bpi_\e$ or
the Galactic priors do separately. 

{More generally, $\bpi_\e$ measurements can provide valuable
information even when they are consistent with zero.  As discussed in the
Appendix to \citet{ob150479}, no ``evidence'' of parallax is required
to introduce the two $\bpi_\e$ parameters because it is known a priori 
that $\pi_\e$ is strictly positive.  Hence, measurements that constrain 
it to be small
provide additional information, even if they are formally consistent with 
zero.}

Another important application of 1-D parallax measurements is that
they can yield a very precise mass estimate when the source and lens
are separately resolved many years after the event \citep{ob03175,gould14}.
That is, such imaging generally yields a very precise measurement of
$\bmu_\rel$, which has the same direction as $\bpi_\e$.  Thus, if 
$\pi_{\e,\parallel}$ is well constrained (and assuming that the $\bpi_\e$
error ellipse is not by chance closely aligned to $\bmu_\rel$),
both $\pi_\e$ and $\theta_\e = \mu_\rel t_\e$ can be precisely determined
from such imaging.  See \citet{ob05071b} and \citet{ob05071c} for 
a case in which
the original 1-D parallax measurement was turned into lens mass and
distance measurements by successively better determinations of $\bmu_\rel$.


{\subsection{{KMT-2019-BLG-0253}
\label{sec:anal-kb190253}}

In the original automated search for anomalies \citep{ob191053}, the
deviation in KMT-019-BLG-0253 from the 1L1S model 
(Figure~\ref{fig:0253lc}) was entirely due to KMTC data forming 
a {0.23 day, roughly linear rising track that lay below the model,
centered on
HJD$^\prime\equiv$ HJD - 2450000 = 8588.81, i.e., $\Delta t_{\rm anom}=-1.8$
days} before the
peak.  The method aggressively removes bad-, or even 
questionable-seeing and high-background 
points, which very frequently produce such ``features''.  Nevertheless,
because the algorithm examines ${\cal O}(10^2)$ nights of data from 
three observatories on ${\cal O}(10^3)$ events, there are numerous such 
single-observatory deviations even after this aggressive automated
culling of the data.
Hence, they must be treated with caution.  However, after the TLC re-reductions,
the previously excluded KMTS points on the same night could be
re-included, and these confirmed the dip, whose duration 
can be estimated $\Delta t_{\rm dip}=0.6\,$days.  Moreover, there are five
OGLE points on this night, which generally track the KMTC deviation,
including three at the beginning of the night when the dip is pronounced.
Finally, dense MOA coverage confirms the subtle ``ridge'' feature
immediately following the dip.
Thus, there is no question that the deviation is of astrophysical
origin.  See Figure~\ref{fig:0253lc}.

{\subsubsection{{Heuristic Analysis}
\label{sec:heuristic-kb190253}}

The 1L1S model yields parameters $t_\e\simeq 60\,$days and $u_0\simeq 0.06$.
{The midpoint of the dip is at} $\Delta t_{\rm anom}= -2.0\,$days, and
$\Delta t_{\rm dip} = 0.6\,$days.  Applying the
heuristic formalism of Section~\ref{sec:anal-preamble}, these yield,
\begin{equation}
\alpha = 61^\circ;
\qquad
s \simeq s_-^\dagger \pm \Delta s;
\qquad
s_-^\dagger = 0.97;
\qquad
q \sim 7\times 10^{-5},
\label{eqn:heur-kb190253}
\end{equation}
where $\pm\Delta s$, which is
induced by the offset of the source trajectory from the caustic,
cannot be
evaluated from the general appearance of the light curve.

{\subsubsection{{Static Analysis}
\label{sec:static-kb190253}}

Notwithstanding this seemingly secure reasoning, we conduct a systematic
search for 2L1S solutions, as described
in Section~\ref{sec:anal-preamble}.
As expected, we find two solutions, which we then seed into
two MCMCs in which all parameters are allowed to vary.  The resulting
parameters are shown in Table~\ref{tab:0253parms}.  
As can be seen from Figure~\ref{fig:0253lc}. 
the two model light curves are virtually indistinguishable.  A classic
case of this
inner/outer degeneracy for the minor image perturbation
is presented by \citet{ob161067}.  However,
in contrast to that case, the ``outer'' solution does not have
a planetary caustic (outside of which the source would pass).  Rather
the planetary and central caustics have merged into a single resonant
caustic (e.g., \citealt{gaudi12}), and 
the ``outer'' solution passes outside the ``planetary wing''
of this resonant caustic.  See Figure~\ref{fig:4caustics}.  Within the context
of the heuristic treatment of Section~\ref{sec:heuristic-kb190253}, 
we find from Table~\ref{tab:0253parms} that the two solutions can be represented
as $(s^\dagger_-\pm \Delta s) = (0.969\pm 0.040)$, while $\alpha=59^\circ$.
Both are in good agreement with Equation~(\ref{eqn:heur-kb190253}).
By contrast, the static-model mass-ratio, $q=4.1\times 10^{-5}$,
is almost a factor two smaller than the heuristic estimate.

We note that the $(s,\alpha)$ geometry, caustic topology, and
inner/outer degeneracy of OGLE-2018-BLG-0677 \citep{ob180677} are the same 
as for KMT-2019-BLG-0253, except that the anomaly comes just after the peak, 
while it comes just before peak for KMT-2019-BLG-0253. In that case,
$(s^\dagger_- \pm \Delta s)=(0.95\pm 0.03)$, which is also
very similar to the present case.

The case of OGLE-2016-BLG-1195 \citep{ob161195a,ob161195b} is
also closely related.  It likewise has an inner/outer degeneracy
with the outer solution having a resonant caustic.  However,
in  that case, the source intersected the binary axis on the
planet's side of the host rather than the opposite side.  Hence,
$s^\dagger_+ >1$, i.e., $(s_+^\dagger \pm \Delta s)= (1.03\pm 0.05)$,
compared to $s_-^\dagger<1$ for KMT-2019-BLG-0253 and OGLE-2018-BLG-0677.
OGLE-2019-BLG-0960 \citep{ob190960} is yet another example of a 
major-image anomaly induced by a similar caustic topology, with 
$(s^\dagger_+ \pm \Delta s)= (1.012\pm 0.015)$.

{\subsubsection{{Parallax Analysis}
\label{sec:parallax-kb190253}}

In the static solution, the event timescale is a substantial fraction
of a year, which implies that it may be possible to measure the 
annual-parallax effect due to Earth's orbital motion.
As mentioned in Section~\ref{sec:anal-preamble}, 
we must at least initially also introduce
the linearized orbital-motion parameters $\bgamma$.
However, we find that the orbital-motion parameters are neither
significantly constrained nor significantly correlated with the
parallax parameters.  Therefore, we suppress these two degrees of 
freedom (dof).  

We initially find a seemingly strongly constrained and
large value of $\pi_\e$.  However, after conducting several tests,
we conclude that this signal is the result of systematics in the
KMT data.  First, by examining the cumulative distribution,
$\Delta\chi^2(t) = \chi^2(t;{\rm static}) - \chi^2(t;{\rm parallax})$,
we find that the great majority of the ``signal'' comes from
late in the light curve, $t>t_0 + 2\,t_\e$.  It is nearly impossible
to arrange a physical situation for which this would be the case.
Moreover, this ``signal'' appears only in the KMT data and not in the
OGLE data.  With fewer data points, we would expect the signal to
be weaker in OGLE, but still present.  Finally, two KMT observatories
(KMTS and KMTA) show
strongly increasing $\Delta\chi^2(t)$ during these late times,
while the third is strongly decreasing.  We also considered that this
``signal'' might be due to ``xallarap'', i.e., orbital motion of the
source about an unseen companion, rather than orbital motion of Earth.
Indeed, there was a stronger xallarap signal than parallax signal.
However, the same inconsistencies between data sets persisted.
And this continued when we fit for parallax and xallarap simultaneously.
We attempted to eliminate only these late-time KMT data, but similar
signatures of systematics remained, albeit at a lower level.  
We are not certain what is the
cause of these systematics, which affect all three KMT telescopes,
but in different directions.  We note that there are several bright
variable stars in the field, though none closer $20^{\prime\prime}$.
These, in principle, could affect KMT photometry at a very low
level, without affecting OGLE photometry, which is overall better.
Then, the very large number of these points could amplify the small
``signal'' in each to create strong trends.

We therefore measure $\bpi_\e$ by fitting only to OGLE data,
but with the planet parameters $(s,q,\alpha,\rho)$ held fixed at
those of the static solution.
The results are shown in Table~\ref{tab:0253parms}.
There are two points to note.  First, the axis ratio of the
error ellipse, $\sigma(\pi_{\e,\perp})/(\sigma(\pi_{\e,\parallel})$,
is large, ranging from 15 to 20, depending on the solution.
That is, this is a 1-D parallax measurement.  Second,
$|\pi_{\e,\parallel}|/\sigma(\pi_{\e,\parallel})\simeq 3$, indicating that
the parallax is robustly detected.  In fact, we find a very similar
value for this parameter from the full data set but with much
smaller errors $\pi_{\e,\parallel} \simeq -0.100\pm 0.009$.  As
$\pi_{\e,\parallel}$ is much less prone to systematics than
$\pi_{\e,\perp}$, we take this as qualitative confirmation of
the correction of the $\pi_{\e,\parallel}$ measurement.

We note that $\rho$ is not
constrained to be different from zero at the $3\,\sigma$ level.
Rather, there is only an upper limit on $\rho$, which
arises because if $\rho$ were sufficiently large, the ``dip'' would
be more rounded than is actually the case.  In fact, this will
prove to be the case for all {six} events presented here.
Nevertheless, this upper limit on $\rho$ 
(so lower limit on $\theta_\e=\theta_*/\rho$) can help constrain
the physical interpretation of the event, as we discuss in
Section~\ref{sec:physical}.

{For completeness, we report the static solutions for the OGLE-only data 
set:
$(\chi^2,t_0,u_0,t_\e,I_S) = 
(1653.7, 8590.5587\pm 0.0134, 0.0558\pm 0.0018, 56.35\pm 1.53, 19.771\pm 0.036)$
and
$(1652.7,8590.5686\pm 0.0144,0.0576\pm 0.0018, 54.91\pm 1.46, 19.735\pm 0.035)$,
for the inner and outer solutions, respectively.  Because there are of
order 9 times fewer OGLE points compared to OGLE+KMT, we expect that the
errors will be of order $\sqrt{9}=3$ times larger than those of the 
static solutions from the full data set, and this expectation is confirmed
by comparison to Table~\ref{tab:0253parms}.  Within these errors,
the static and parallax solutions are consistent for these parameters.

Comparing the OGLE-only static and parallax fits, we see that the
errors in the latter are of order two times larger for $(u_0,t_\e,I_S)$.
This is partially explained by correlations between each of these three 
parameters and $\pi_{\e,\perp}$.  That is, to leading order, $\pi_{\e,\parallel}$ and
$\pi_{\e,\perp}$ give rise to light-curve distortions that are,
respectively, anti-symmetric and symmetric in $(t-t_0)$
\citep{smp03,gould04}.  Because the \citet{pac86} formula is even
in $(t-t_0)$ for $(u_0,t_\e,I_S)$, these three parameters are correlated with
$\pi_{\e,\perp}$, which has a large uncertainty.  
We note that this same effect is present in the three other
events with parallax solutions that are analyzed below, although it is
strongest in the case of KMT-2019-BLG-0253.
}

{\subsection{{OGLE-2018-BLG-0506}
\label{sec:anal-ob180506}}

Figure~\ref{fig:0506lc} 
shows a clear $\Delta I\sim 0.06$ mag dip in the KMTC data at
$\Delta t_{\rm anom} \sim 0.54\,$days after the peak, with
a duration $\Delta t_{\rm dip}= 0.4\,$days.  The depression 
defined by these 26 data points, which are taken in good seeing 
($1.2^{\prime\prime}$--$1.6^{\prime\prime}$) and low background is confirmed
by the two contemporaneous OGLE points.  Hence, the anomaly is secure.

{\subsubsection{{Heuristic Analysis}
\label{sec:heuristic-ob180506}}

The 1L1S model yields parameters $t_\e\simeq 24\,$days and $u_0\simeq 0.09$.
As noted above, $\Delta t_{\rm anom}=+0.54\,$days and
$\Delta t_{\rm dip}=0.4\,$days.  Applying the
heuristic formalism of Section~\ref{sec:anal-preamble}, these yield,
\begin{equation}
\alpha = 104^\circ;
\qquad
s \simeq s_-^\dagger \pm \Delta s;
\qquad
s_-^\dagger = 0.95; 
\qquad
q \sim 17\times 10^{-5} .
\label{eqn:heur-ob180506}
\end{equation}
Again, $\pm\Delta s$,  which is induced by
the offset of the source trajectory from the caustic,
cannot be evaluated from the general appearance of the light curve.

{\subsubsection{{Static Analysis}
\label{sec:static-ob180506}}

The grid search described in Section~\ref{sec:anal-preamble} returns
two local minima, which we then further refine by allowing all
parameters to vary in the MCMC.  The resulting solutions are
given in Table~\ref{tab:0506parms} 
and illustrated in Figure~\ref{fig:0506lc}.  
The two solutions can be represented
as $(s^\dagger_-\pm \Delta s) = (0.961\pm 0.099)$, while $\alpha=106^\circ$,
both in good agreement with Equation~(\ref{eqn:heur-ob180506}).
The mass ratios of the two solutions are 
$q_{\rm inner} = (7.8\pm 2.3)\times 10^{-5}$ and
$q_{\rm outer} = (7.6\pm 2.3)\times 10^{-5}$, respectively.
Note that these measurements
are in only qualitative agreement with the estimate in
Equation~(\ref{eqn:heur-ob180506}).

Figure~\ref{fig:4caustics}
shows that the caustic topologies are identical to the case of
KMT-2019-BLG-0253.  That is, for the inner solution, the source
{traverses} the trough that extends along the minor-image axis between
the central and planetary caustics, while in the outer solution,
these two triangular caustics have merged into a resonant caustic.

{\subsubsection{{Parallax Analysis}
\label{sec:parallax-ob180506}}

The timescale of this event is relatively short, and the source is faint.
Therefore, we do not expect to be able to measure the
microlens parallax $\bpi_\e$.  Nevertheless, as a matter of due diligence,
we explore this possibility.  We first find after including both
$\bpi_\e$ and the lens orbital parameters $\bgamma$, that the latter
are not meaningfully constrained (relative to the physical condition 
$\beta<0.8$).
Hence, we suppress these two degrees of freedom.  
{As in the case of KMT-2019-BLG-0253 correlated errors in the KMT
baseline data induced spurious parallax signals, but in contrast to
that case, there was no evidence of spurious signals in the body of
this event.  Hence, most KMT baseline data were removed, i.e., roughly
70\% of the full-year KMT data set.  For consistency, the same cuts
were applied for the static fits.}
The results are given
in Table~\ref{tab:0506parms}.  These show that both
$\pi_{\e,\parallel}$ and $\pi_{\e,\perp}$ are consistent with zero at the
$1\,\sigma$ level, which is reflected by the fact that $\chi^2$ is 
essentially unchanged by the addition of two degrees of freedom.
Nevertheless, the fact that $\pi_{\e,\parallel}$ is constrained to be
near zero within relatively small errors can be a significant constraint
in the Bayesian analysis of Section~\ref{sec:physical}.  Moreover, 
none of the other parameters are
significantly affected by the inclusion of $\bpi_\e$ in the fit.  We 
therefore adopt the parallax solutions for our final result.

{\subsection{{OGLE-2018-BLG-0516}
\label{sec:anal-ob180516}}

Again, there is clear dip in the light curve, which has 
$\Delta t_{\rm anom}=2.2\,$days and duration
$\Delta t_{\rm dip}=0.5\,$days.
It is most clearly defined by KMTC data, but these are supported
by contemporaneous OGLE data.  Moreover, the beginning of the dip
is traced by KMTS data, while the end of the dip is supported by both
MOA and KMTA data.
Hence, the anomaly is secure.

{\subsubsection{{Heuristic Analysis}
\label{sec:heuristic-ob180516}}

The 1L1S model yields parameters $t_\e\simeq 25\,$days and $u_0\simeq 0.10$.
As noted above, $\Delta t_{\rm anom}=2.2\,$days and $\Delta t_{\rm dip}=0.5\,$days.
Applying the
heuristic formalism of Section~\ref{sec:anal-preamble}, these yield,
\begin{equation}
\alpha = 131^\circ;
\qquad
s \simeq s_-^\dagger \pm \Delta s;
\qquad
s_-^\dagger = 0.93;
\qquad
q \sim 10\times 10^{-5}.
\label{eqn:heur-ob180516}
\end{equation}
Again, $\pm\Delta s$ cannot be determined at this level of analysis.

{\subsubsection{{Static Analysis}
\label{sec:static-ob180516}}

The grid search described in Section~\ref{sec:anal-preamble} returns
two local minima, which we then further refine by allowing all
parameters to vary in the MCMC.  The resulting solutions are
given in Table~\ref{tab:0516parms} and illustrated in Figure~\ref{fig:0516lc}.  
The two solutions can be represented
as $(s^\dagger_-\pm \Delta s) = (0.936\pm 0.069)$, while $\alpha=130^\circ$,
in good agreement with Equation~(\ref{eqn:heur-ob180516}).
The mass ratios of the two solutions are 
$q_{\rm inner} = (12.9\pm 1.4)\times 10^{-5}$ and
$q_{\rm outer} = (13.2\pm 1.4)\times 10^{-5}$, respectively, which are
also in good agreement with Equation~(\ref{eqn:heur-ob180516}).

Figure~\ref{fig:4caustics}
shows that the caustic topologies are again identical to the cases of
KMT-2019-BLG-0253 and OGLE-2018-BLG-0506.

{\subsubsection{{Parallax Analysis}
\label{sec:parallax-ob180516}}

The {overall situation here is similar} to that of 
OGLE-2018-BLG-0506
(Section~\ref{sec:parallax-ob180506}).  
{However, in contrast to that case, there were
no spurious signals from the KMT baseline data, so the full data set
was included in the fit.}
Again we test for, then
suppress the orbital-motion parameters $\bgamma$.  
In this case, we find that
$\pi_{\e,\parallel}$ is consistent with zero at the $\simeq 1.5\,\sigma$ level,
so the $\chi^2$ improvement is $\simeq 2$ for 2 degrees of freedom.
Again, we find little difference for the other parameters
between the static and parallax analyses.  And again, we adopt the
parallax solution for our final result, noting that the $\pi_{\e,\parallel}$
constraint can play some role in the Bayesian analysis of 
Section~\ref{sec:physical}.

{\subsection{{OGLE-2019-BLG-1492}
\label{sec:anal-ob191492}}

Again, there is clear dip in the light curve, which has 
$\Delta t_{\rm anom}=+1.6\,$days and duration
$\Delta t_{\rm dip}=0.8\,$days.
It is most clearly defined by KMTC data, but these are supported
by contemporaneous OGLE data.  Moreover, there are ``ridges''
around the dip, which are supported by all observatories.

{\subsubsection{{Heuristic Analysis}
\label{sec:heuristic-ob191492}}

The 1L1S model yields parameters $t_\e\simeq 50\,$days and $u_0\simeq 0.05$.
As noted above, $\Delta t_{\rm anom}=1.6\,$days and 
$\Delta t_{\rm dip}=0.8\,$days.
Applying the
heuristic formalism of Section~\ref{sec:anal-preamble}, these yield,
\begin{equation}
\alpha = 123^\circ;
\qquad
s \simeq s_-^\dagger \pm \Delta s;
\qquad
s_-^\dagger = 0.97;
\qquad
q \sim 18\times 10^{-5}.
\label{eqn:heur-ob191492}
\end{equation}
Again $\pm\Delta s$ cannot be determined at this level of analysis.

{\subsubsection{{Static Analysis}
\label{sec:static-ob191492}}

The grid search described in Section~\ref{sec:anal-preamble} returns
two local minima, which we then further refine by allowing all
parameters to vary in the MCMC.  The resulting solutions are
given in Table~\ref{tab:1492parms} and illustrated in Figure~\ref{fig:1492lc}.  
The two solutions can be represented
as $(s^\dagger_-\pm \Delta s) = (0.971\pm 0.073)$, while $\alpha=122^\circ$,
in good agreement with Equation~(\ref{eqn:heur-ob191492}).
The mass ratios of the two solutions are 
$q_{\rm inner} = (19.1\pm 5.8)\times 10^{-5}$ and
$q_{\rm outer} = (17.6\pm 5.4)\times 10^{-5}$, respectively, which are
in good agreement with Equation~(\ref{eqn:heur-ob191492}).

Figure~\ref{fig:4caustics}
shows that the caustic topologies are again identical to the cases of
KMT-2019-BLG-0253, OGLE-2018-BLG-0506, and OGLE-2018-BLG-0516.

{\subsubsection{{Parallax Analysis}
\label{sec:parallax-ob191492}}

The situation here is {similar to that of OGLE-2018-BLG-0506 (Section
\ref{sec:parallax-ob180516}).  Spurious parallax signals from the KMT baseline
force us to remove most of these data points, corresponding to
about half of the full-year KMT data set.}
Again we test for, then
suppress the orbital-motion parameters $\bgamma$.  Again we find
that $\bpi_\e$ is only weakly constrained (and only in the $\pi_{\e,\parallel}$
direction).  Again, we find little difference for the other parameters
between the static and parallax analyses.  And again, we adopt the
parallax solution for our final result.

{\subsection{{OGLE-2018-BLG-0977}
\label{sec:anal-ob180977}}

The KMTS data show a strong and sudden dip ending in a flat
minimum, which is characteristic of the source passing into the
trough between the two triangular minor-image caustics, with the
entrance close to one of these caustics.  One therefore expects
the source to pass over the other caustic, an expectation that
is confirmed by the rapid drop of the KMTA data during the first three points
of the night, which span 69 minutes.  Thus, KMTS and KMTA give independent,
and completely consistent evidence for a minor-image caustic crossing.

The center of the dip occurs $\Delta t_{\rm anom} = +4.0\,$days after peak.
Because the source crosses the two triangular caustics, the
$\Delta t_{\rm dip}=0.5\,$day depression is a much more clearly defined
structure than in the four previous cases.

{\subsubsection{{Heuristic Analysis}
\label{sec:heuristic-ob180977}}

The 1L1S model yields parameters $t_\e\simeq 20\,$days and $u_0\simeq 0.15$.
As noted above, 
$\Delta t_{\rm anom}= +4.0\,$days and $\Delta t_{\rm dip}= 0.5\,$days.  Applying the
heuristic formalism of Section~\ref{sec:anal-preamble}, these yield,
\begin{equation}
\alpha = 143^\circ;
\qquad
s = s_-^\dagger = 0.88;
\qquad
q \simeq 5\times 10^{-5}.
\label{eqn:heur-ob180977}
\end{equation}
That is, because the source crosses the planetary caustic, there
is only one solution.  In such cases, we expect the mass-ratio estimate
to be quite accurate.

{\subsubsection{{Static Analysis}
\label{sec:static-ob180977}}

The grid search described in Section~\ref{sec:anal-preamble} returns
only one local minimum, which we then further refine by allowing all
parameters to vary in the MCMC.  
{We find that, due to the extreme faintness of the source $I_S\sim 21$
and shortness of the event ($t_\e=20\,$days), correlations in the KMT
baseline data impact even the static fit.  We therefore eliminate these
data, keeping only $8260<{\rm HJD}^\prime < 8296$}.
The resulting solution is
given in Table~\ref{tab:0977parms} and illustrated in Figure~\ref{fig:0977lc}.
In particular, $(s,q,\alpha) = (0.879,4.15\times 10^{-5},144^\circ)$
in good agreement with Equation~(\ref{eqn:heur-ob180977}).

Figure~\ref{fig:4caustics} shows that, as expected, and in contrast to the
previous four events, the source crosses the caustic.

{\subsubsection{{Parallax Analysis}
\label{sec:parallax-ob180977}}

When we incorporate $\bpi_\e$ {and} $\bgamma$ into the fits, we
find that neither is meaningfully constrained.  Therefore, we
adopt the parameters of the static model.

{
{\subsection{{OGLE-2019-BLG-0953}
\label{sec:anal-ob190953}}}

In contrast to the previous five events, the amplitude of the
anomaly (a dip at HJD$^\prime\simeq 8643.7$, 
i.e., $\Delta t_{\rm anom}=+5.4\,$days) is not much larger
than the individual KMT error bars.  See Figure~\ref{fig:0953lc}.
However, this event lies in a small
$(0.4\,{\rm deg}^2)$ zone that is covered by four fields and so
is observed at $\Gamma = 8\,{\rm hr}^{-1}$.  All four fields from
two observatories (KMTC and KMTS) participate in the same overall
trend during the night of the dip.  Moreover the dip is confirmed
by eight OGLE points.  Hence, it is secure.

{\subsubsection{{Heuristic Analysis}
\label{sec:heuristic-ob190953}}}

The 1L1S model yields parameters $t_\e\simeq 11.5\,$days and $u_0\simeq 0.5$.
We estimate $\Delta t_{\rm dip}=0.9\,$days, and, as noted above, 
$\Delta t_{\rm anom}=5.4\,$days.
Applying the
heuristic formalism of Section~\ref{sec:anal-preamble}, these yield,
\begin{equation}
\alpha = 137^\circ;
\qquad
s \simeq s_-^\dagger \pm \Delta s;
\qquad
s_-^\dagger = 0.71;
\qquad
q \sim 17\times 10^{-5}.
\label{eqn:heur-ob190953}
\end{equation}
Again $\pm\Delta s$ cannot be determined at this level of analysis.

{\subsubsection{{Static Analysis}
\label{sec:static-ob190953}}}

In sharp contrast to the other four non-caustic crossing events
(all the above except OGLE-2018-BLG-0977),
the grid search described in Section~\ref{sec:anal-preamble} returns
only one local minimum.  We then further refine this by allowing all
parameters to vary in the MCMC.  The resulting solution is given in 
Table~\ref{tab:0953parms} and illustrated in Figure~\ref{fig:0953lc}.  

The physical reason that the inner-outer degeneracy is broken
in this case can be assessed by comparing the caustic topology of
KMT-2019-BLG-0953 to the other four non-caustic-crossing cases in
Figure~\ref{fig:4caustics}.  For this event, the (unique)
solution has the ``outer'' geometry, but with isolated planetary
caustics, whereas in the other four cases, the outer solution had
a resonant caustic.  Hence, if there were two solutions, they
would lie on opposite sides of isolated planetary caustics.
This alternate topology follows from the fact that $u_{\rm anom}\sim 0.7$,
compared to 0.07--0.13 for the other four cases, which means that it is
far from the resonant regime (for low-$q$ companions).  Because
minor-image isolated caustics have substantial structure in the 
direction perpendicular to the binary axis, the inner-outer degeneracy
can usually be broken when both trajectories pass isolated caustics,
unless the trajectories are themselves close to perpendicular,
as in the case of OGLE-2016-BLG-1067 \citep{ob161067}.

Because there is only one solution, $s^\dagger$ is expected to be
merely offset from
the (unique) value of $s$, rather than being nearly equal to the
average of two values, as in the other four cases.  Nevertheless,
this offset is small: $s - s^\dagger = 0.03$, while 
the fitted values of both $q$ and $\alpha$ are all in good agreement
with Equation~(\ref{eqn:heur-ob190953}).

Note that there is only a weak constraint on the source size, $\rho$, 
which corresponds to $t_* < 0.33\,$day.

{\subsubsection{{Parallax Analysis}
\label{sec:parallax-ob190953}}}

Similarly to several other cases, we find that the baseline data
would generate a spurious parallax signal.  We therefore restrict
the analysis to $8600 < t < 8670$, i.e., roughly $t_0\pm 3\,t_\e$,
which we then also apply to the static solution.
After doing so, we
find that the microlens parallax $\bpi_\e$ is not meaningfully
constrained, and we therefore adopt the static solution.
}

{\section{{Source Properties}
\label{sec:cmd}}


Overall, our principal objective in this section is to measure 
the source-star angular radius, $\theta_*$,
and then to either make measurements of, or obtain lower limits upon,
\begin{equation}
\theta_\e = {\theta_*\over\rho};\
\qquad
\mu_\rel = {\theta_\e\over t_\e}.
\label{eqn:thetae_murel}
\end{equation}
depending on whether $\rho$ is measured, or itself only has an upper limit.
For the first step, we use the method of \citet{ob03262}.  That is,
we first measure the color-magnitude offset between the source and
the centroid of the red clump on a color-magnitude diagram (CMD),
\begin{equation}
\Delta[(V-I),I] = [(V-I),I]_s - [(V-I),I]_{\rm cl}.
\label{eqn:deltacmd}
\end{equation}
Next we adopt $(V-I)_{\rm cl,0}=1.06$ from \citet{bensby13}, and we evaluate
$I_{\rm cl,0}$ from Table~1 of \citet{nataf13}, based on the Galactic
longitude of the event.  We then find the dereddened color and magnitude
of the source
\begin{equation}
[(V-I),I]_{s,0} = [(V-I),I]_{\rm cl,0} + \Delta[(V-I),I].
\label{eqn:cmd0}
\end{equation}
We then apply the $VIK$ color-color relations of \citet{bb88} to transform from
$V/I$ to $V/K$, and finally, we apply the color/surface-brightness relations
of \citet{kervella04} to obtain $\theta_*$.  After propagating the
measurement errors, we add 5\% to the error in quadrature to take account
of systematic effects in the method as a whole.

Figure~\ref{fig:allcmd} shows the CMDs for all {six} events.  
For one of these (OGLE-2018-BLG-0516), the CMD field stars are
drawn from the pyDIA \citep{pydia} 
reduction of a $2^\prime\times 2^\prime$ square centered
on the event.  The source color is determined from regression of
the $V$-band and $I$-band source fluxes during the event.
The source magnitude is calculated from the value in Table~\ref{tab:0516parms}
added to the offset between the pyDIA and pySIS systems as determined
by regression of their respective light curves.

For the remaining four events
(KMT-2019-BLG-0253,
OGLE-2018-BLG-0516,
OGLE-2019-BLG-1492, and
OGLE-2018-BLG-0977), the CMD field stars are drawn from
the calibrated OGLE-III catalog \citep{oiiicat}, with
respective radii of 
$80^{\prime\prime}$,
$120^{\prime\prime}$,
$100^{\prime\prime}$, and
$90^{\prime\prime}$.  These values were chosen by balancing the competing
demands of attaining sufficient density to determine the position
of the clump and minimizing the effects of differential extinction.
To determine the source position on the OGLE-III CMD, the same
steps were first taken as for OGLE-2018-BLG-0516.  Then the
pyDIA CMD was calibrated to OGLE-III, allowing the source position
to be transformed.

Table~\ref{tab:cmd} shows the measurements and logical train
described above for all {six} events.  In addition, we comment
on specific aspects of particular events, below.

{\subsection{{KMT-2019-BLG-0253}}
\label{sec:cmd-kb190253}}

KMT-2019-BLG-0253 has significant blended light, which raises the 
question of whether or not this blend could be due to the lens.
The CMD position of the baseline-object 
$[(V-I),I]_{\rm base} = (3.30,18.84)\pm (0.18,0.06)$ was determined
by transforming the KMTC02 pyDIA field-star photometry at the position
of the event to the OGLE-III system\footnote{Note that the CMD position
of the baseline object is below the threshold of detection 
(roughly $V\sim 21.5$) on the OGLE-III CMD, but still has only moderate
color error in the KMT field-star photometry.  This is because the $V$-band
photometry is carried out by different methods.  For OGLE-III, sources
on the $V$ and $I$ templates are found independently and then matched based
on astrometry.  For KMT sources are identified only in $I$ band.  Then
the $V$-band measurement is derived from the $V$-band flux at this
$I$-band location, averaged over all available images.}.   
We also measured the SDSS $i$ flux at this position from seven images taken with
the Canada-France-Hawaii Telescope (CFHT) in $0.5^{\prime\prime}$ seeing,
transformed to the OGLE-III system.  This confirmed the KMTC02
$I_{\rm base}$ measurement within errors.  Subtracting the source flux from
the baseline flux yields the blended flux
$[(V-I),I]_b = (3.67,19.42)\pm (0.46,0.10)$.
The baseline object and blended light are shown in Figure~\ref{fig:allcmd}
as green and magenta points, respectively.

We transform the source position (as determined from KMTC02 difference images)
to the CFHT $0.5^{\prime\prime}$ images and measure its offset from the baseline
object, finding,
\begin{equation}
\Delta \btheta(E,N) = \btheta_{\rm base} - \btheta_x = (-31,-6)\pm(7,10)\,\mas .
\label{eqn:dtheta}
\end{equation}
This offset is formally inconsistent with zero at more than $4\,\sigma$,
which would nominally rule out the lens as the origin of the blended light.
In fact, as we show below, there are physical effects that can account
for this offset, even if the blended light is due to the lens.

However, first we note that 
this small offset can be compared with the overall surface
density of field stars that are brighter than the blend, $I_b=19.42$,
which we find {from the OGLE CMD} 
to be $N=0.012\,{\rm arcsec}^{-2}$.  Taking 
Equation~(\ref{eqn:dtheta}) at face value and noting that the blend:source
flux ratio is 1.75, the separation of the source and blend is 50 mas.
Thus, the probability of such a chance superposition is 
$p=\pi N((11/7)\Delta\theta)^2\sim 1\times 10^{-4}$,
which is to say, virtually ruled out.  Hence, the blended light is due
either to the lens itself, a companion to the lens, or a companion to the 
source.  

One possibility for the offset in Equation~(\ref{eqn:dtheta}), just mentioned,
is that it is due to a companion to the source or the lens.  For the
source (at $D_S\sim 8\,\kpc$), the projected separation would correspond to
$\sim 250\,\au$.  For the lens, it would be proportionately closer, e.g.,
$\sim 150\,\au$ at $D_L\sim 5\,\kpc$.  
Such $\log(P/{\rm day})\sim 6$ stellar companions
are among the most common.  See Figure~7 of \citet{dm91}.  On the other
hand, if the blend were a companion to the source, it would be a subgiant,
i.e., of almost identical mass to the source.  From Table~7 of 
\citet{dm91}, this would be quite rare.  It is more plausible that
the blend is a companion to the lens, simply because there are no
independent constraints on its mass ratio or distance.

However, another possibility is that an ambient star 
(contributing some, but not all, of the blended light)
is corrupting the measurement of $\btheta_{\rm base}$.  Suppose, for example, that
10\% of the baseline flux were due to an ambient star at $\sim 300\,\mas$
from the source (and lens).  This would be too faint and too close to be
separately resolved, even in 500 mas seeing.  Therefore, the
fitting program would find the flux centroid displaced by 
$0.1\times 300\,\mas= 30\,\mas$, which is what is observed.
We can estimate the surface density of stars that are 
{$2.5\pm0.5$ mag fainter than the baseline object (i.e., $M_I=4.5\pm 0.5$)}
using the \citet{holtzman98} luminosity
function, based on {\it Hubble Space Telescope (HST)} observations of
Baade's Window.  We must first multiply by a factor 3 based on the
relative surface density of clump stars in the KMT-2019-BLG-0253 field
compared to Baade's Window (\citealt{nataf13}, D.Nataf 2019, private
communication).  We then find a surface 
{density $N=1.0\,{\rm arcsec}^{-2}$.}
Hence, the probability that one such star falls 
{within $0.3^{\prime\prime}$  of the lens is about 25\%.}

We conclude that the blend is most likely either the lens itself or
a companion to the lens.  It could be a companion to the source with
much lower probability.  And it is very unlikely to be an ambient star.

{\subsection{{OGLE-2018-BLG-0506}}
\label{sec:cmd-ob180506}}

We note that the pyDIA reductions of the 
KMTC data are only approximately calibrated.  However, this has no
practical importance because the filters are near standard
and the only quantity entering the calculations that follow is the
offset between the source and the clump.  Moreover,
this offset is small.  See Figure~\ref{fig:allcmd}.
For reference, we note that in the other other four events,
the KMT$\rightarrow$OGLE-III correction was found to be
$[(V-I),I]_{\rm s, kmt} - [(V-I),I]_{\rm s, ogle-iii} =$
[0.06, 0.22],
[0.17, 0.12],
[0.14, 0.01], and 
[0.03 ,0.22], for
KMT-2019-BLG-0253,
OGLE-2018-BLG-0516,
OGLE-2019-BLG-1492, and
OGLE-2018-BLG-0977, respectively.

We find a best estimate for the blended light $I_b=21.08$ on the 
OGLE-IV system and $I_b=21.11$ on the KMTC41 pyDIA system.  Even ignoring
the uncertainties in this estimate due to the mottled background light,
which can be considerable (e.g., \citealt{kb180029}), this measurement
would play very little role when treated as an upper limit on
lens light.  For example, even at $M=1\,M_\odot$ (the $1\,\sigma$ upper
limit derived in Section~\ref{sec:physical}), and at $D_L = 5.6\,\kpc$,
the lens magnitude would be $I_L \sim 22.2$.  
Therefore, we do not pursue a detailed investigation of this blend.

{\subsection{{OGLE-2018-BLG-0516}}
\label{sec:cmd-ob180516}}

The KMTC01 pyDIA analysis formally yields a negative blended
flux that, after conversion to the OGLE-III system, corresponds to
an $I_b=21.1$ ``anti-star''.  This level of negative blending would
be easily explained by errors in the PSF-fitting photometry of faint
stars in the baseline images, 
even without taking account of the unresolved mottled background.
Hence, it is of no concern in itself, and can be regarded as consistent
with zero.  The field has extinction $A_I=1.87$.  Hence, 
a lens with $M=0.84\,M_\odot$ (the $1\,\sigma$ upper limit derived in
Section~\ref{sec:physical}) at $D_L\sim 7.0\,\kpc$ would have brightness
$I_L = 20.8$.  This is comparable to the error in the measurement
of the blended flux (as demonstrated by its negative value).  Hence,
no useful limits on the lens flux can be derived.

{\subsection{{OGLE-2019-BLG-1492}}
\label{sec:cmd-ob191492}}

For {four of the six} events, the position of the source on
the CMD corresponds to what is expected for a reasonably
common class of bulge star.  These are
turnoff stars (KMT-2019-BLG-0253 and OGLE-2018-BLG-0516),
clump giant (OGLE-2018-BLG-0506), and
mid-G dwarf (OGLE-2018-BLG-0977).  However, the source position
for OGLE-2019-BLG-1492 is somewhat puzzling and deserves
further investigation.

The source has a similar color to the Sun, but is
about 1.45 mag fainter than the Sun would be if it were at the
mean distance of the clump.  This could, in principle, be resolved
by it being 1.45 mag dimmer than the Sun or 1.45 mag more  distant
than the clump, or some combination.  For example, for the first,
it could be that the source is a relatively unevolved metal-poor star.
Another potential explanation is that the source color has not
been correctly measured.  For example, if the source were 0.2
mag redder, it would be expected to be 1 mag or more dimmer than
the Sun.  However, the color is determined from regression of
about a dozen well-magnified $V$-band points that lie along the regression line.
Hence this explanation appears unlikely.  Regardless of the
exact explanation of the discrepancy, it has little practical
effect.  That is, even if $\theta_*$ were double the value
that we have estimated in Table~\ref{tab:cmd}, the constraints provided by 
the resulting limits on $\theta_\e$ and $\mu_\rel$ 
would still be too weak to matter.

The blended light has $I_b=19.47$ on the calibrated OGLE-III scale.
However, the baseline $V$-band photometry is not precise enough
to reliably determine $(V-I)_b$.  In principle, this blended light
might be due primarily to the lens.  However, we find from CFHT baseline
images in $0.45^{\prime\prime}$ seeing that the blend is displaced
from the event by $0.33^{\prime\prime}$.  Therefore, it cannot be the lens,
and so must be either a companion to the lens or to the source, or an
ambient star that is unrelated to the event.  We find that surface density
of stars that are brighter than the blend $(I<I_b)$ is 
$0.33\,{\rm arcsec}^{-2}$.  Thus, the probability of an ambient star within
a circle of radius equal to the separation is 11\%.  This is 
the most likely explanation for the blended light.

{\subsection{{OGLE-2018-BLG-0977}}
\label{sec:cmd-ob180977}}

In Table~\ref{tab:cmd}, we only list a $3\,\sigma$ upper
limit on $\rho$ because the light curve is formally consistent with
$\rho=0$ at this level.  However, it should be noted
that at the $1\,\sigma$ and $2\,\sigma$
levels, $\rho$ is constrained by
$\rho=1.9^{+0.3}_{-0.6}\times 10^{-3}$  and
$\rho=1.9^{+0.6}_{-1.6}\times 10^{-3}$, respectively.  These tighter limits
on $\rho$ result primarily from the source crossing a caustic.
When, evaluating the physical parameters in Section~\ref{sec:physical},
we will make full use of the $\chi^2$ fit values as a function of $\rho$.
The blended light is a relatively bright star on the foreground 
main sequence
$[(V-I),I]_{b,OGLE-III} = (17.39,1.98)$, which raises the possibility
that this is the lens.  However, we find from examining the CFHT baseline
image that this blended light is a resolved star with a separation 
of $\sim 0.5^{\prime\prime}$.  Thus, it is definitely not the lens, nor
can it be a companion to the source.  This leaves two possibilities:
companion to the lens or ambient star.  
The surface density of foreground main-sequence stars brighter than the
blend is $N=0.0085\,{\rm arcsec}^{-2}$.  Hence, the probability of
such an ambient star within $0.5^{\prime\prime}$ is $p=1/150$.
This is relatively low, but the prior probability that a given star
has a substantially more massive companion at separation
$a_\perp\ga$1000--2000 au (for distance 2--4 kpc) is also of order 1\%
or less.  Thus, no information about the lens system can be inferred
from the presence of this bright blend.

{
{\subsection{{KMT-2019-BLG-0953}}
\label{sec:cmd-kb190953}}

There was significant difficulty measuring the source color of 
KMT-2019-BLG-0953.  As with the other four events that lie in the 
OGLE-III catalog \citep{oiiicat} footprint, our approach is to first
measure the source color in the KMT system by regression and then to
transform this color to the OGLE-III system by comparison of field stars.
The results for all events are shown as blue points in Figure~\ref{fig:allcmd}.
However, for KMT-2019-BLG-0953, we applied this procedure five times,
using the independent data sets from KMTC02, KMTC42, KMTC03, KMTC43, and KMTS02.
Strikingly, these five determinations are comprised of two groups:
(KMTC02, KMTC43) were closely grouped near $(V-I)_{s,\rm OGLE-III} \simeq 1.94$, 
while
(KMTC03, KMTC42, KMTS02) were closely grouped near 
$(V-I)_{s,\rm OGLE-III} \simeq 2.05$, The errors in each case were near 0.04.
One approach would simply be to take the average of these
five measurements, which would yield a mean and standard error of the mean
of $2.006\pm 0.018$.  This joint fit has $\chi^2=9.1$ for four degrees of
freedom, which has a Gaussian probability of 
$p = (1  + \chi^2/2)\exp(-\chi^2/2) = 6\%$.  Under normal circumstances,
this value would not be low enough to raise serious concerns.

However, there are two further issues that must be taken into account.
First, using the clump centroid from Table~\ref{tab:cmd},
this would lead to a dereddened source color $(V-I)_{s,0} = 0.51\pm 0.02$.
Such blue stars are extremely rare in the bulge.  Given that the source is
2.3 magnitudes below the clump, it would have to be either a blue straggler
or a relatively young star.  Alternatively it could be an F dwarf in the
disk, relatively close to the Galactic bulge.  This would also be 
unexpected.

The second issue is that the color measurement may be impacted by the
presence of a bright neighbor at $1.0^{\prime\prime}$, which is about 2 mag
brighter than
the source.  DIA works by first subtracting the reference image convolved
with a kernel and then multiplying the resulting difference image by
the point spread function (PSF).  If the subtraction were perfect, the
only impact of the neighbor would be to add photon noise, which is
already taken into account in the measurement process.  However, the 
PSF varies over the field, particularly near the edge of the field (recall that
this event lies in the overlap of two neighboring fields), so that
the PSF (and resulting kernel) are inevitably less than perfect.
Then, residuals from an imperfectly subtracted, bright, overlapping neighbor
can impact the photometry.  Thus, the difficulties induced by this neighbor 
might plausibly account for the relatively high $\chi^2/$dof of the joint 
fit, in which case the high scatter would be due to systematics, so that
the errors could not be treated as Gaussian.

We resolve these challenges as follows.  First, we derive the color
from the three redder measurements and ignore the two bluer measurements
as plausibly due to systematics.  Second, we adopt the standard deviations
of these measurements, rather then the standard error of the mean,
in order to reflect the non-Gaussian errors.
Third, with these assumptions, the source is very plausibly a bulge
star, and so we assign it bulge kinematics in the Bayesian analysis of
Section~\ref{sec:physical}.

However, we also briefly discuss the implications for both present and
future work, if the source is later proved by followup observations to
be bluer than we have assumed.

First, such a bluer source would yield a smaller $\theta_*$.
In particular, it would be $\sim 10\%$ smaller if the source were
0.11 mag bluer.  Under normal circumstances, this would affect the
Bayesian estimates by reducing both $\theta_\e$ and $\mu_\rel$ by 10\%.
However, in the present case, we have only weak lower limits on these
quantities (see Table~\ref{tab:cmd}), so there is essentially no impact.
Furthermore, if the lens and source are separately resolved by future adaptive 
optics (AO) observations, then $\mu_\rel$ (and so $\theta_\e = \mu_\rel t_\e$)
will be directly determined from these observations, so that the current
$\theta_*$ estimate will become moot.

The main impact would be that a blue source would almost certainly imply
that the source is in the disk, which then implies that the lens
(which must lie in front of the source) is also in the disk.  Thus, adopting
a blue color would affect the Bayesian estimates.  The error bars of these
estimates are already quite large, and we regard the probability of a blue
source to be sufficiently small to ignore this possibility.  Hence,
the main role of this section will be to alert workers carrying out
future followup observations to the difficulties of the color measurement,
so that they can properly take these into account in their analyses.

}

{\section{{Physical Parameters}
\label{sec:physical}}

When both $\theta_\e$ and $\pi_\e$ are measured, one can directly
determine the lens mass and lens-source relative parallax by inverting
Equation~(\ref{eqn:pie}), and one can then infer the lens distance $D_L$
by estimating the source parallax $\pi_S$, which also yields 
an estimate of the projected planet-host separation, $a_\perp$
\begin{equation}
M={\theta_\e\over\kappa\pi_\e};
\qquad
D_L = {\au\over \theta_\e\pi_\e + \pi_S};
\qquad
a_\perp = {s\over \pi_\e + \pi_S/\theta_\e}\au.
\label{eqn:pie2}
\end{equation}

Unfortunately, we do not measure both $\theta_\e$ and $\pi_\e$ for
any of the {six} events.  For four of the {six} events, we obtain
a constraint on $\bpi_\e$ (a ``one-dimensional parallax'' measurement),
while for the {other two (OGLE-2018-BLG-0977 and KMT-2019-BLG-0953)},
we do not obtain any constraint of $\bpi_\e$.
For each of the 
{six} events, we obtain only a constraint on $\rho$ (and so on
$\theta_\e=\theta_*/\rho$).  We therefore must estimate the physical
parameters from a Bayesian analysis using Galactic-model priors.

We apply the procedures of \citet{ob191053} to each of the 
{six} planetary events.
In each case, we consider an ensemble of simulated events drawn
from a Galactic model.  We weight these events by the product of
five terms,
\begin{equation}
\Gamma\propto \theta_\e\mu_\rel\chi^2(t_\e)\chi^2(\bpi_\e)\chi^2(\rho).
\label{eqn:gamma}
\end{equation}
The first two terms are simply the cross section and speed from
the very general rate formula ``$\Gamma = n\sigma v$''.  The third
term reflects how well the Einstein timescale of the simulated event
matches that of the actual event, within the latter's measurement
error.  Similarly, the fourth term reflects the match of $\bpi_\e$
between the simulated and observed events.  The only difference
is that, because $\bpi_\e$ is a two-dimensional (2-D) quantity, the
comparison is made via a 2-D error ellipse rather than a 1-D error bar.
{For OGLE-2018-BLG-0977 and KMT-2019-BLG-0953, 
the best value of $t_\e$ and its error bar
come from Table~\ref{tab:0977parms}.  For the remaining four events,
the parallax values and error ellipses come from 
$(\pi_{\e,\parallel},\pi_{\e,\perp},\psi)$ in Tables 
\ref{tab:0253parms}--\ref{tab:1492parms}, while the $t_\e$ and error bars
come from the corresponding solutions.}

The only element of Equation~(\ref{eqn:gamma}) that is somewhat less
familiar is $\chi^2(\rho)$.  Normally, one measures $\rho$ and infers
$\theta_\e = \theta_*/\rho$ (with some error bar) and so this term
is usually {expressed in} the form $\chi^2(\theta_\e)$.  However,
lacking $\rho$ measurements, we proceed as follows.

From the MCMC, we evaluate the lower envelope of $\chi^2$ as a function
of $\rho$.  For example, for the case of KMT-2019-BLG-0253, this is similar
to a Heaviside step function, e.g., $\chi^2(\rho) = 100\Theta(0.0035)$,
in which any value $\rho<0.0035$ would be permitted with no $\chi^2$
penalty, and all others would be strictly excluded.  By contrast,
for the case of OGLE-2018-BLG-0977, $\chi^2(\rho)$ has a clear minimum
at $\rho\sim 0.0019$, i.e., well below the $3\,\sigma$ limit $\rho<0.0031$.
However, even $\rho=0$ is formally disfavored at only the $\sim 2\,\sigma$
level.

For each simulated event, $i$, we evaluate $\rho_i =\theta_*/\theta_{\e,i}$
where $\theta_*$ is the value determined in Section~\ref{sec:cmd} and
$\theta_{\e,i}$ is the Einstein radius of the simulated event.

For all the events {except OGLE-2018-BLG-0977 and KMT-2019-BLG-0953}, 
there are multiple
solutions.  For these cases, we combine the results from the different
solutions, with the simulated events from each solution weighted
by the $\Delta\chi^2$ (as given in 
Tables~\ref{tab:0253parms}--\ref{tab:0953parms})
for that solution.
Table~\ref{tab:phys}  shows the results.
For more details on the Bayesian method, see \citet{ob191053}

{In Section~\ref{sec:cmd-kb190253}, we assessed that blended
light in KMT-2019-BLG-0253 (magenta point in Figure~\ref{fig:allcmd})
was most likely due to the lens itself or a companion to the lens.
In either case, it would be at the distance to the lens, which,
from Table~\ref{tab:phys}, corresponds to distance modulus
$(m-M)\simeq 13.45\pm 0.75$.  The blend lies $2.47\pm 0.10$ mag below
the clump, which has $I_{\rm cl,0}=14.44$.  At $D_L\sim 5\,\kpc$, the
lens is almost certainly behind essentially all the dust.  Thus, if the
blend is either the lens or its companion, then its absolute
magnitude is $M_{I,\rm bl}=3.46\pm 0.75$.  This range is somewhat too
luminous for a middle K dwarf, which is indicated by the blend's
dereddened color being somewhat redder than the clump.  On the
other hand a K dwarf would be quite consistent with the mass
range shown in Table~\ref{tab:phys}.  One possibility that would
account for all the constraints would be that the lens is part of
a roughly equal-mass K-dwarf binary.  Alternatively, the lens could
be somewhat closer than indicated by the $1\,\sigma$ Bayesian range,
in which case it would be intrinsically 
dimmer (at the same observed brightness).  These issues can only be 
resolved by AO followup observations.  In many
cases, fruitful AO observations must wait for the lens and source
to move apart sufficiently to permit them to be separately resolved,
and for this event the time required for such separation
is highly uncertain because there is only a lower limit on $\mu_\rel$.
However, in the present case, immediate AO observations would be highly
informative because the blend is brighter than the source, and the
observations would allow one to confirm (or rule out) the conjecture
that the astrometric offset observed from the ground was due to
low-level contamination by an ambient star.
}

}

{In Section~\ref{sec:anal-preamble}, we argued that 1-D parallax
measurements can play a role in constraining the Bayesian estimates
of the physical parameters, even when the $\chi^2$ improvement is small
(or even zero).  We illustrate that impact in the present case by
repeating the Bayesian analysis without the parallax constraint and
reporting the resulting mass estimates for the first four events in
Table~\ref{tab:phys}:
$M_{\rm host} = (0.63_{-0.34}^{+0.38},0.53_{-0.30}^{+0.41},0.42_{-0.26}^{+0.40},
0.58_{-0.33}^{+0.40})M_\odot$.  Thus, by including the parallax constraint,
the mass estimates were changed by
$(+11,+19,+12,+21)\%$, which is notable but below $1\,\sigma$ in all cases.
This is not surprising given that the fractional Bayesian ranges
$\delta_M \equiv (M_{+1\,\sigma} - M_{-1\,\sigma})/2 M_{\rm median}$ 
in Table~\ref{tab:phys} are relatively large: 
$\delta_{M,\rm parallax} = (0.46,0.56,0.78,0.50)$.  However, another important
impact is that these fractional ranges are themselves reduced by
including the parallax constraint:
$\delta_{M,\rm parallax}/\delta_{M,\rm static}  = (0.81,0.84,0.99,0.79)$.
That is, the fractional range was substantially reduced in all but
one case.
}


{\section{{Discussion}
\label{sec:discuss}}

We have applied the method of Paper I \citep{ob191053} to the six
KMT prime fields (BLG01/02/03/41/42/43) for the 2018-2019 seasons,
with the aim of obtaining a homogeneously selected sample of
planets with $q<2\times 10^{-4}$.  Specifically, we obtained TLC reductions
of all events that had $q<3\times 10^{-4}$ based on pipeline reductions,
and we thoroughly investigated these events using the
TLC reductions.  While a full analysis
must await the further application of this method to more seasons
(at least 2016-2017) and more fields, it is of considerable interest
to review the properties of this initial sample.

Table~\ref{tab:2018recover} 
shows all
the previously known planets that were recovered by our approach,
regardless of $q$.
Table~\ref{tab:dchi20} shows the {11} events with planets $q<2\times 10^{-4}$
({7} discovered and 4 recovered) by our approach, rank ordered by
$\Delta\chi^2_0$ of anomaly detection (see Paper I).  A striking feature
of this table is that all the new discoveries are at the top, i.e., they
all have lower $\Delta\chi^2_0$ than any of the recoveries.  This implies
that, broadly speaking, machine selection of anomaly candidates 
is more robust than by-eye selection.  Of course, it is still necessary
for humans to vet these candidates, whether initially selected by machine
or by eye.

In fact, there are two exceptions to this pattern that are not
reflected in Table~\ref{tab:dchi20}.  First,
OGLE-2018-BLG-0677Lb \citep{ob180677}, with $q=9.1\times 10^{-5}$,
was not found by our approach because it was below our
$\Delta\chi^2_0 = 75$ threshold, yet it was found by eye.  Second,
KMT-2018-BLG-1025Lb \citep{kb181025}, with $q_{\rm best} =8\times 10^{-5}$,
is not shown in Table~\ref{tab:dchi20} because it has a second solution
at $q =16\times 10^{-5}$, with $\Delta\chi^2=8.4$, and so it is not suitable
for investigating the mass-ratio function.  Nevertheless, because
it was recovered at $\Delta\chi^2_0=140$, it would break the simple
pattern of Table~\ref{tab:dchi20} if it were included.  Both examples show that
humans are competitive with machines in some individual cases.
Nevertheless, the overall message from Table~\ref{tab:dchi20} remains the same:
many subtle planetary signatures escape recognition in by-eye searches.
Note also from Table~\ref{tab:dchi20} that $\Delta\chi^2_0$ is hardly
correlated with $q$, if at all.

{\subsection{{Eleven Discovered/Recovered $q<2\times 10^{-4}$ Planets}
\label{sec:6planets}}

Our approach yielded a total of {11} planets with $q<2\times 10^{-4}$ in
the KMT prime fields during 2018-2019, including OGLE-2019-BLG-1053Lb,
which was reported by \citet{ob191053}, the {six} reported here, and
four previously discovered planets\footnote{Not including
KMT-2018-BLG-1025Lb \citep{kb181025}, which,
as mentioned above, is excluded from the sample because
it has two degenerate solutions that are well-separated in $q$.
},
OGLE-2018-BLG-0532Lb \citep{ob180532},
OGLE-2018-BLG-0596Lb \citep{ob180596},
OGLE-2018-BLG-1185Lb \citep{ob181185}, and
KMT-2019-BLG-0842Lb \citep{kb190842}.  

There are {four} known $q<2\times 10^{-4}$ planets from 2018-2019 that 
were not found in this search:
KMT-2018-BLG-0029Lb \citep{kb180029},
OGLE-2019-BLG-0960Lb \citep{ob190960}, 
{KMT-2018-BLG-1988Lb \citep{kb181988}}, and
OGLE-2018-BLG-0677Lb \citep{ob180677}.  The first {three} of these
do not lie in prime fields, while the last failed our 
$\Delta\chi^2_0>75$ cut in the initial automated anomaly search.
{Note that KMT-2018-BLG-1988Lb has a very large (factor 2) $1\,\sigma$
error in $q$, so that it will almost certainly not be included in
mass-ratio studies.}

Finally, although it is not directly germane to the present study,
we note that our approach failed to recover (or fully recover) three
known planets with $q>2\times 10^{-4}$.  
The first is KMT-2019-BLG-1715LABb \citep{kb191715} which is a 3L2S
event.  The automated AnomalyFinder identified the anomaly
generated by the $q_2=0.25$ binary companion,
which dominated
(and so suppressed recognition of) the planetary companion
$q_3=4\times 10^{-3}$.  The AnomalyFinder would have to be modified to
take account of the possibility of planet-binary systems to have
detected this planet.  The second case is
that of KMT-2019-BLG-1953Lb \citep{kb191953}, which is a 
$q\sim 2\times 10^{-3}$ ``buried planet'' \citep{mb07400},
i.e., its signature is submerged in the finite-source effects near the
peak of a high-magnification event, $A_\max\sim 1000$.  It was
identified in our search as a finite-source-point-lens (FSPL) anomaly.  
In principle, all such events should
be systematically searched for planets because of their overall 
strong sensitivity, particularly to Jovian mass-ratio planets.
However, this stage was not pursued in the present context because of the
weak sensitivity of such events
to very low-$q$ planets, which is the focus of the present effort.

The third case is OGLE-2018-BLG-1011Lb,c \citep{ob181011}.  
This is a two-planet system
for which the AnomalyFinder reports a single anomaly that contains
both planets.  However, in contrast to the case of the binary+planet
system KMT-2019-BLG-1715LABb that was mentioned above, this is
not really a shortcoming.  That is, the AnomalyFinder identifies
the principal anomaly, but does not itself classify the anomaly
as ``planetary'' or ``binary''.  Rather, the operator must make the
decision that an anomaly is ``potentially planetary''.  All such
events must be thoroughly investigated
before they can be published, and such detailed investigations, which
are carried out using TLC reductions, are far more sensitive to
multiplicity of planets than any potential machine search based
on pipeline reductions.  By contrast, the detection of a binary
system will not, of itself, trigger such a detailed investigation.
Because OGLE-2018-BLG-1011 was a well-known planetary system at the
time that the AnomalyFinder was run on 2018 data, no investigation
was needed.  However, it would have been triggered if the planets
were not already known.

For completeness, we note that all three of these planetary events, i.e.,
OGLE-2018-BLG-1011, KMT-2019-BLG-1715, and KMT-2019-BLG-1953, were
designated as ``anomalous'' by the AnomalyFinder.
However, because there
were multiple anomalies in the three cases (either a third body or
finite source effects that dominated over the planet), the
AnomalyFinder did not identify all of the {\it planetary} anomalies.

The cumulative mass-ratio distribution of these {11} planets
is presented in Figure~\ref{fig:cum}, which shows
that they span the range
$-5<\log q < -3.7$.  We have not yet measured the efficiency
of our selection, but the automatic selection should be less
sensitive to lower-$q$ planets.  

In particular, the 
sensitivity $\xi(q)$ of any wide-angle 
ground-based survey that depends primarily on a $\chi^2$ threshold to
detect planets will be approximately a power law, which gradually
steepens toward lower $q$.  First, in the regime where the size of the
sources (dominated by upper-main-sequence and turnoff stars) is smaller
than the caustic size, the cross section for the source to interact
with the caustics, or with the magnification structures that extend from
them, scales as either $q^{1/2}$ or $q^{1/3}$ for resonant and planetary
caustics, respectively.  Second, as $q$ falls, the duration of the
anomaly declines according to the same power laws, which for fixed
source brightness $I_s$, proportionately reduces $\chi^2$.  Hence,
for otherwise similar events, the $\chi^2$ threshold restricts lower-$q$
planets to the brighter end of the luminosity function. Thus, if the luminosity
function were a strict power law (as it roughly is for $0.5<M_I<3$,
Figure~5 of \citealt{holtzman98}), then the efficiency would
also be a power law, $\xi(q)\propto q^\gamma$.  In fact, the 
luminosity becomes much shallower for $M_I\ga 3$, which makes $\xi(q)$
become gradually shallower toward higher $q$.  Finally, for low mass ratios
($\log q \la -4$) where turnoff sources become comparable to or larger
than the caustic size, two new effects take hold.  The first is that 
the cross section of interaction is fixed by the source size, rather 
than declining with $q$ as the caustic size.  By itself, this would
make the slope shallower toward lower $q$.  However, the second effect
is that the flux deviation due to the caustic is washed out by the extended
source, which at fixed $I_s$ reduces $\chi^2$ and so restricts detections
to the brighter end of the luminosity function, i.e., more so than
for the case analyzed above of small sources.  The net effect is
to maintain a steepening $\xi(q)$ in this regime.  A corollary of this logic
is that if one overestimates (or underestimates) the source sizes in the
efficiency calculation, then one will derive a $\xi(q)$ that is too steep
(or too shallow) in the $\log q\la -4$ regime.  However
Figure~7 of \citet{suzuki16} shows that this is a relatively modest
effect.

Therefore,
while no rigorous conclusions can yet be drawn, 
a declining sensitivity function suggests that the low mass-ratio
planets in this homogeneously selected sample represents a substantial
underlying population. This raises questions about the previously
inferred paucity of planets at
low $q$ \citep{suzuki16,kb170165}.

We note that it is already known that the low-$q$
2018-2019 planets 
KMT-2018-BLG-0029Lb ($\log q=-4.74$) and
OGLE-2019-BLG-0960Lb ($\log q=-4.90$) will be recovered when
we apply our approach to the non-prime KMT fields.  
See Figure~\ref{fig:cum}.  While we cannot include these 
in the homogeneously selected distribution because we do not
yet know what new planets will be discovered in the non-prime
fields, these detections prove that such
planets exist. So, it will not be the case that only larger mass-ratio
planets are found in {these fields}.  Therefore,
it would require many such discoveries with $\log q\ga -4.3$
to maintain a strong case for a mass-ratio function that
declines with declining $q$ in the $q<2\times 10^{-4}$ regime.
We briefly discuss the reasons that we do not consider this to be likely in 
Section~\ref{sec:sens-gamma}.

{\subsection{{Angular Distribution}
\label{sec:angular}}

An interesting difference between the recovered versus discovered 
low-$q$ planets is that the former all have source trajectories
that are closer to the planet-host axis than the latter.
The two groups have angular offsets
($6^\circ$, $8^\circ$, $15^\circ$, $27^\circ$), and
($31^\circ$, $36^\circ$, $45^\circ$, $50^\circ$, $58^\circ$, $59^\circ$, $74^\circ$),
respectively.  In particular, the two $q<10^{-4}$ recovered planets
have trajectories very close to the planet-host axis
($6^\circ$ and $8^\circ$).
Although we have excluded KMT-2018-BLG-1025, we note that if the
favored $q<10^{-4}$ solution is indeed correct, then it also has a source
trajectory close to the binary axis ($5^\circ$).  This probably means
that it is much easier to spot by eye the extended anomalies
due to such oblique encounters, compared to the short dips or
bumps in the events containing the newly discovered planets}.  
This higher sensitivity 
of oblique trajectories was already noted by \citet{kb190842},
\citet{ob190960}, and \citet{kb200414}.

{\subsection{{Sensitivity as a Function of $\Gamma$}
\label{sec:sens-gamma}}

It is overall less likely that the {seven} newly discovered 
(as opposed to recovered) planets would 
have been discovered if exactly the same event had occurred in
lower cadence fields.  All {seven planets} were {\it newly} discovered 
because a 
short-lived, low-amplitude perturbation was densely covered in the prime
fields.  Their short duration and low amplitude severely 
diminished\footnote{We say ``diminished'' rather than ``prevented'' because
OGLE-2018-BLG-0677 \citep{ob180677} was recognized by eye despite the
fact that it was below the threshold of our machine search.} the chance 
that they would be noticed in by-eye searches, while their dense coverage
enabled relatively high $\Delta\chi^2_0$ during the small, short deviation.
The same event would likely still escape by-eye detection 
in a lower-cadence field,
while the $\Delta\chi^2_0$ would be lower by the same factor as the cadence.
Four of the {seven} discovered,
planets had nominal cadences of $\Gamma=4\,{\rm hr}^{-1}$,
two (OGLE-2018-BLG-0516 and OGLE-2018-BLG-0977) 
had $\Gamma=2\,{\rm hr}^{-1}$, and
one (KMT-2019-BLG-0953 had $\Gamma=8\,{\rm hr}^{-1}$.  If we
simply scale $\Delta\chi^2_0$ by $\Gamma$, then for KMTNet's seven 
$\Gamma=1\,{\rm hr}^{-1}$ fields, the {seven} 
$q<2\times 10^{4}$ discoveries listed
in Table~\ref{tab:dchi20} would have
$\Delta\chi^2_0 = (22,56,41,98,65,35,75)$.  For these lower-cadence
fields, the detection threshold will be set to
$\Delta\chi^2_0 = 50$ (rather than 75) because the problem of contamination
by low-level systematics is substantially reduced.  Hence, four of the 
{seven}
planetary events would have been investigated, and plausibly published.
On the other hand, for
KMTNet's 11 $\Gamma=0.4\,{\rm hr}^{-1}$ fields, the corresponding
numbers are $\Delta\chi^2_0 = (9,22,16,39,26,14,30)$, 
which are clearly hopeless.
The same applies, ipso facto, to KMTNet's 
three $\Gamma=0.2\,{\rm hr}^{-1}$ fields.

{Regarding higher mass-ratio planets, $q>2\times 10^{-4}$, these are
generally much easier to identify by eye.  Our preliminary review of
the higher mass-ratio sample from the high-cadence fields confirms
this assessment.  The ratio of recovered-to-discovered planets that 
unambiguously have mass ratios $q<2<10^{-4}$ is 4:7.
For comparison there are 16 recovered planets with $2\times 10^{-4}<q<0.03$
in Table~\ref{tab:2018recover}.  If the recovered-to-discovered ratio
were the same as for low-$q$ planets, then we would expect 28 high-$q$
discoveries.  While we have not yet completed our investigation, we are
confident that the actual number of high-$q$ discoveries will be well below
this number.
}

Thus, we expect that when our method is applied to lower-cadence fields,
it will generally recover the by-eye detected events, but 
the fraction of new discoveries will likely be smaller.

{\subsection{{Conclusions}
\label{sec:conclude}}

{
The KMTNet AnomalyFinder was applied to 2018-2019 prime-field data
and returned 11 low mass-ratio ($q<2\times 10^{-4}$) planets, of which
four were previously discovered by eye, one was previously reported in
Paper I \citep{ob191053}, and {six} are reported here for the first time.
This 7:4 ratio suggests that many low mass-ratio planets are missed
in by-eye searches.  By contrast there was only one by-eye discovery
that was missed by AnomalyFinder, and this because it was below the
$\chi^2$ threshold of the algorithm.

While the investigation of the higher mass-ratio ($q>2\times 10^{-4}$) detections
is not yet complete, a preliminary review suggests that by-eye searches
are much more effective in this regime.  

All six of the newly discovered low-$q$ planets have higher angular offsets 
($31^\circ<\alpha^\prime<74^\circ$) relative to the planet-host axis compared
to the four previous discoveries ($6^\circ<\alpha^\prime<27^\circ$), which suggests
that the ``dilated'' anomalies induced by acute trajectories are easier
to discover by eye in the low-$q$ regime \citep{kb190842}.

Combined, these results suggest that low mass-ratio planets may be more
common than previously believed.  However, systematic application
of the AnomalyFinder to the lower-cadence fields and to additional seasons,
combined with an efficiency analysis currently underway 
(Y.K.~Jung et al., in prep), will be required to make firm statements
about this possibility.

Of the six planets reported in this paper, one could benefit from immediate
AO observations.  We found that the centroid of the
blended light in KMT-2019-BLG-0253 lies within about 50 mas of the source
and is most likely dominated by the lens or a companion to the lens.
AO observations could greatly clarify or resolve the nature of this blend.
}

\acknowledgments 
This research has made use of the KMTNet system operated by the Korea
Astronomy and Space Science Institute (KASI) and the data were obtained at
three host sites of CTIO in Chile, SAAO in South Africa, and SSO in
Australia.
Work by C.H. was supported by the grants of National Research Foundation
of Korea (2019R1A2C2085965 and 2020R1A4A2002885).
The OGLE project has received funding from the National Science
Centre, Poland, grant MAESTRO 2014/14/A/ST9/00121 to AU.
The MOA project is supported by JSPS KAKENHI Grant Number JSPS24253004, JSPS26247023, JSPS23340064, JSPS15H00781, and JP16H06287.
W.Z., S.M., X.Z. and H.Y. acknowledge support by the National Science Foundation of China (Grant No. 11821303 and 11761131004). This research uses data obtained through the Telescope Access Program (TAP), which has been funded by the TAP member institutes.
We are very grateful to the instrumentation and operations teams at CFHT who fixed several failures of MegaCam in the shortest time possible, allowing its return onto the telescope and these crucial observations.

 \begin{deluxetable}{lrrrrrr}
 \tablecolumns{7} \tablewidth{0pc}
 \tablecaption{\textsc{Event Names, Cadences, Alerts, and Locations}}
 \tablehead{\colhead{Name} & 
\colhead{$\Gamma\,({\rm hr}^{-1})$} &
\colhead{Alert Date} &
\colhead{RA$_{\rm J2000}$} &
\colhead{Dec$_{\rm J2000}$} &
\colhead{$l$} &
\colhead{$b$} }
 \startdata
KMT-2019-BLG-0253 & 4.0 & 2 Apr 2019 & 17:51:31.82 & $-29$:33:55.7 & +0.13 & $-1.43$\\
OGLE-2019-BLG-0410& 1.0 \\
MOA-2019-BLG-127 &  4.0 \\
\hline
OGLE-2018-BLG-0506 & 4.0 & 30 Mar 2018 & 17:50:31.16 & $-31$:55:26.6 & $-2.01$ & $-2.45$ \\
KMT-2018-BLG-0835  & 0.3 \\
\hline
OGLE-2018-BLG-0516 & 0.3 & 1 Apr 2018 & 17:58:33.63 & $-31$:15:44.6 & $-0.57$ & $-3.59$ \\
MOA-2018-BLG-107   & 1.0 \\
KMT-2018-BLG-0808  & 2.0 \\
\hline
OGLE-2019-BLG-1492 & 3.0 &  6 Oct 2019 & 18:00:23.15 & $-28$:37:52.1 & +1.91 & $-2.63$ \\
KMT-2019-BLG-3004  & 4.0 \\
\hline
OGLE-2018-BLG-0977 & 0.8 & 3 Jun 2018 & 17:54:01.47 & $-30$:36:17.3 & $-0.49$ & $-2.42$ \\
KMT-2018-BLG-0728  & 2.0 \\
\hline
KMT-2019-BLG-0953 & 8.0 &27 May 2019 & 17:57:21.26 & $-28$:40:28.7 & $+1.54$ & $-2.08$ \\

 \enddata
 \label{tab:names}
 \end{deluxetable}

\begin{table}[htb]
    \renewcommand\arraystretch{1.05}
    \centering
    \caption{2L1S Parameters for KMT-2019-BLG-0253}
    \begin{tabular}{c|c c|c c c c}
    \hline
    \hline
    Parameters &  \multicolumn{2}{c|}{Static} & \multicolumn{4}{c}{Parallax} \\
      & Inner & Outer & Inner $u_0 > 0$ & Inner $u_0 < 0$ & Outer $u_0 > 0$ & Outer $u_0 < 0$ \\
    \hline
    $\chi^2$  & 14661.2 & 14660.9 & 1640.0 & 1640.2 & 1642.0 & 1641.9 \\
    dof      &  14661   & 14661   & 1640   & 1640   & 1640   & 1640  \\
    \hline
    $t_{0}$ & 8590.5774 & 8590.5814 & 8590.5618 & 8590.5552 & 8590.5645 & 8590.5599 \\
              & 0.0053 &    0.0055 &    0.0183 &    0.0162 &    0.0172 &    0.0154 \\
    $u_{0}$  & 0.0555 & 0.0559 & 0.0538 & $-$0.0532 & 0.0565 & $-$0.0566 \\
            & 0.0006 & 0.0006 & 0.0034 &    0.0035 & 0.0026 &    0.0027 \\
    $t_\e$ (days) & 57.01 & 56.66 & 58.77 & 59.60 & 56.08 & 55.76 \\
                  & 0.55 &  0.52 &  4.13 &  4.13 &  2.74 &  2.80 \\
    $s$  & 0.9289 & 1.0092 & ... & ... & ... & ... \\
         & 0.0074 & 0.0088 & ... & ... & ... & ... \\
    $q$ ($10^{-5}$) & 4.07 & 4.10 & ... & ... & ... & ... \\
                   & 0.75 & 0.72 & ... & ... & ... & ... \\
    $\alpha$ (rad) & 1.0230 & 1.0223 & ... & ... & ... & ... \\
                   & 0.0043 & 0.0041 & ... & ... & ... & ... \\
    $\rho$         & $<$0.0045 & $<$0.0045 & ... & ... & ... & ...  \\
  $\pi_{\rm E,N}$       &...&...& $ -0.550$ & $  0.509$ & $ -0.362$ & $  0.198$ \\
                        &...&...& $  0.609$ & $  0.490$ & $  0.482$ & $  0.445$ \\
  $\pi_{\rm E,E}$       &...&...& $  0.028$ & $  0.114$ & $  0.053$ & $  0.102$ \\
                        &...&...& $  0.074$ & $  0.038$ & $  0.062$ & $  0.036$ \\
  $\pi_{\e,\parallel}$  &...&...& $ -0.089$ & $ -0.089$ & $ -0.094$ & $ -0.092$ \\
                        &...&...& $  0.030$ & $  0.030$ & $  0.029$ & $  0.029$ \\
  $\pi_{\e,\perp}$      &...&...& $ -0.544$ & $  0.514$ & $ -0.354$ & $  0.203$ \\
                        &...&...& $  0.613$ & $  0.491$ & $  0.485$ & $  0.445$ \\
  $\psi$ (deg)          &...&...& $  276.4$ & $  272.8$ & $  276.5$ & $  272.8$ \\
    $I_{\rm S, KMTC02}$ & 19.780 & 19.771 & 19.665 & 19.680 & 19.611 & 19.607 \\
                     &  0.012 &  0.012 &  0.071 &  0.074 &  0.051 &  0.054 \\
    \hline
    \hline
    \end{tabular}
    \tablecomments{For parallax solutions, $(s,q,\rho,\alpha)$ are held fixed at
their static values, and only the OGLE data are included in the fit.
      The upper limits on $\rho$ are at $3\sigma$.}
    \label{tab:0253parms}

\end{table}

\begin{table}[htb]
    \renewcommand\arraystretch{1.05}
    \centering
    \caption{2L1S parameters for OGLE-2018-BLG-0506}
    \begin{tabular}{c|c c|c c c c}
    \hline
    \hline
    Parameters &  \multicolumn{2}{c|}{Static} & \multicolumn{4}{c}{Parallax} \\
      & Inner & Outer & Inner $u_0 > 0$ & Inner $u_0 < 0$ & Outer $u_0 > 0$ & Outer $u_0 < 0$ \\
    \hline
    $\chi^2$  & 2998.2 & 2997.8 & 2996.0 & 2995.9 & 2995.7 & 2995.5 \\
    dof       & 2998   & 2998   & 2996   & 2996   & 2996   & 2996 \\
    \hline
    $t_{0}$ 
& 8224.1580 & 8224.1584 & 8224.1577 & 8224.1577 & 8224.1581 & 8224.1580 \\
&    0.0026 &    0.0028 &    0.0046 &    0.0042 &    0.0046 &    0.0041 \\
    $u_{0}$  & 0.0884 & 0.0884 & 0.0878 & $-$0.0878 & 0.0879 & $-$0.0878 \\
            & 0.0012 & 0.0012 & 0.0014 &    0.0013 & 0.0013 &    0.0014 \\
    $t_\e$ (days) & 23.86 & 23.87 & 23.84 & 23.81 & 23.85 & 23.84 \\
                   & 0.28 &  0.28 &  0.35 &  0.33 &  0.35 &  0.37 \\
    $s$  & 0.8612 & 1.0594 & 0.8564 & 0.8548 & 1.0673 & 1.0653 \\
         & 0.0176 & 0.0214 & 0.0153 & 0.0148 & 0.0188 & 0.0189 \\
    $q$ ($10^{-5}$) & 7.78 & 7.63 & 8.22 & 8.48 & 8.46 & 8.16 \\
                   & 2.26 & 2.34 & 2.33 & 2.32 & 2.35 & 2.33 \\
   $\alpha$ (rad) & 1.8536 & 1.8532 & 1.8546 & $-$1.8541 & 1.8544 & $-$1.8554 \\
                  & 0.0065 & 0.0067 & 0.0077 &    0.0084 & 0.0078 &    0.0080 \\
    $\rho$ & $<$0.012 & $<$0.014 & $<$0.012 & $<$0.012 & $<$0.014 & $<$0.014 \\
  $\pi_{\rm E,N}$       &...&...& $  0.621$ & $ -0.794$ & $  0.495$ & $ -0.752$ \\
                        &...&...& $  1.636$ & $  1.637$ & $  1.645$ & $  1.683$ \\
  $\pi_{\rm E,E}$       &...&...& $  0.018$ & $ -0.094$ & $  0.006$ & $ -0.090$ \\
                        &...&...& $  0.145$ & $  0.125$ & $  0.147$ & $  0.128$ \\
  $\pi_{\e,\parallel}$  &...&...& $  0.035$ & $  0.036$ & $  0.037$ & $  0.035$ \\
                        &...&...& $  0.034$ & $  0.034$ & $  0.035$ & $  0.034$ \\
  $\pi_{\e,\perp}$      &...&...& $  0.620$ & $ -0.799$ & $  0.494$ & $ -0.757$ \\
                        &...&...& $  1.642$ & $  1.641$ & $  1.652$ & $  1.687$ \\
  $\psi$ (deg)          &...&...& $  274.9$ & $  274.2$ & $  275.0$ & $  274.2$ \\
    $I_{\rm S, KMTC01}$ & 19.163 & 19.163 & 19.170 & 19.169 & 19.169 & 19.170 \\
                    &  0.015 &  0.015 &  0.017 & 0.017 &  0.016 &  0.017 \\
    \hline
    \hline
    \end{tabular}
    \tablecomments{
The upper limits on $\rho$ are at $3\sigma$.}
    \label{tab:0506parms}
\end{table}

\begin{table}[htb]
    \renewcommand\arraystretch{1.05}
    \centering
    \caption{2L1S parameters for OGLE-2018-BLG-0516}
    \begin{tabular}{c|c c|c c c c}
    \hline
    \hline
    Parameters &  \multicolumn{2}{c|}{Static} & \multicolumn{4}{c}{Parallax} \\
      & Inner & Outer & Inner $u_0 > 0$ & Inner $u_0 < 0$ & Outer $u_0 > 0$ & Outer $u_0 < 0$ \\
    \hline
    $\chi^2$  & 5074.4 & 5067.7 & 5072.4 & 5072.7 & 5065.7 & 5065.8 \\
    dof       & 5068  &  5068   & 5066   & 5066   & 5066   & 5065 \\
    \hline
    $t_{0}$ & 8227.6483 & 8227.6514 & 8227.6515 & 8227.6513 & 8227.6541 & 8227.6542 \\
            &    0.0055 &   0.0054 &    0.0057 &    0.0058 &    0.0061 &    0.0057 \\
    $u_{0}$  & 0.1042 & 0.1050 & 0.1034 & $-$0.1034 & 0.1044 & $-$0.1042 \\
              & 0.0013 & 0.0014 & 0.0015 & 0.0016 & 0.0015 & 0.0015 \\
    $t_\e$ (days) & 24.96 & 24.85 & 25.47 & 25.12 & 25.06 & 24.97 \\
                  &  0.25 &  0.25 &  0.39 &  0.39 &  0.37 &  0.36 \\
    $s$  & 0.8674 & 1.0055 & 0.8679 & 0.8676 & 1.0059 & 1.0063 \\
         & 0.0041 & 0.0048 & 0.0042 & 0.0043 & 0.0047 & 0.0048 \\
    $q$ ($10^{-5}$) & 12.90 & 13.18 & 12.80 & 12.92 & 12.99 & 13.13 \\
                   &  1.35 &  1.40 &  1.34 &  1.44 &  1.46 &  1.42 \\
    $\alpha$(rad) & 2.2775 & 2.2754 & 2.2733 & $-$2.2764 & 2.2718 & $-$2.2749 \\
                  & 0.0039 & 0.0037 & 0.0087 &    0.0093 & 0.0077 &    0.0081 \\
    $\rho$ & $<$0.011 & $<$0.011 & $<$0.011 & $<$0.011 & $<$0.011 & $<$0.011 \\
  $\pi_{\rm E,N}$       &...&...& $ -0.461$ & $  0.078$ & $ -0.510$ & $  0.070$ \\
                        &...&...& $  1.049$ & $  1.094$ & $  0.938$ & $  0.999$ \\
  $\pi_{\rm E,E}$       &...&...& $ -0.108$ & $ -0.054$ & $ -0.111$ & $ -0.053$ \\
                        &...&...& $  0.113$ & $  0.092$ & $  0.105$ & $  0.084$ \\
  $\pi_{\e,\parallel}$  &...&...& $  0.062$ & $  0.060$ & $  0.058$ & $  0.058$ \\
                        &...&...& $  0.041$ & $  0.039$ & $  0.041$ & $  0.041$ \\
  $\pi_{\e,\perp}$      &...&...& $ -0.469$ & $  0.074$ & $ -0.519$ & $  0.066$ \\
                        &...&...& $  1.054$ & $  1.097$ & $  0.943$ & $  1.002$ \\
  $\psi$ (deg)          &...&...& $  275.7$ & $  274.4$ & $  275.9$ & $  274.2$ \\
    $I_{\rm S, KMTC01}$ & 19.420 & 19.411 & 19.429 & 19.429 & 19.419 & 19.420 \\
                    &  0.014 &  0.015 &  0.017 &  0.018 &  0.016 &  0.017 \\
    \hline
    \hline
    \end{tabular}
    \tablecomments{
The upper limits on $\rho$ are at $3\sigma$.}
    \label{tab:0516parms}
\end{table}

\begin{table}[htb]
    \renewcommand\arraystretch{1.05}
    \centering
    \caption{2L1S parameters for OGLE-2019-BLG-1492}
    \begin{tabular}{c|c c|c c c c}
    \hline
    \hline
    Parameters &  \multicolumn{2}{c|}{Static} & \multicolumn{4}{c}{Parallax} \\
      & Inner & Outer & Inner $u_0 > 0$ & Inner $u_0 < 0$ & Outer $u_0 > 0$ & Outer $u_0 < 0$ \\
    \hline
   $\chi^2$  & 6792. & 6793.4 & 6789.0 & 6789.6 & 6790.2 & 6790.6 \\
   dof       & 6791  & 6791   & 6789   & 6789   & 6789   & 6789 \\
    \hline
    $t_{0}$ & 8763.137 & 8763.142 & 8763.120 & 8763.125 & 8763.131 & 8763.129 \\
           &    0.027 &    0.026 &    0.032 &    0.032 &    0.032 &    0.033 \\
    $u_{0}$  & 0.0507 & 0.0508 & 0.0473 & $-$0.0480 & 0.0480 & $-$0.0478 \\
            & 0.0032 & 0.0031 & 0.0043 &     0.0042 & 0.0045 &   0.0042 \\
    $t_\e$ (days) & 50.1 & 50.1 & 52.8 & 52.4 & 52.4 & 52.1 \\
                  &  2.7 &  2.7 &  4.2 &  4.0 &  4.1 &  3.9 \\
    $s$  & 0.898 & 1.044 & 0.903 & 0.904 & 1.043 & 1.045 \\
         & 0.015 & 0.018 & 0.014 & 0.015 & 0.018 & 0.019 \\
    $q$ ($10^{-5}$) & 19.1 & 17.6 & 18.0 & 17.8 & 17.3 & 17.6 \\
                    & 5.8 &  5.4 &  4.9 &  5.4 &  4.9 &  5.0 \\
    $\alpha$ (rad) & 2.123 & 2.120 & 2.140 & $-$2.135 & 2.134 & $-$2.139 \\
                   & 0.023 & 0.022 & 0.026 &    0.030 & 0.030 &    0.031 \\
    $\rho$ & $<$0.009 & $<$0.009 & $<$0.009 & $<$0.009 & $<$0.009 & $<$0.009 \\
  $\pi_{\rm E,N}$       &...&...& $ -0.342$ & $  0.143$ & $ -0.210$ & $  0.575$ \\
                        &...&...& $  0.858$ & $  0.957$ & $  0.922$ & $  0.815$ \\
  $\pi_{\rm E,E}$       &...&...& $ -0.062$ & $ -0.026$ & $ -0.042$ & $ -0.001$ \\
                        &...&...& $  0.093$ & $  0.102$ & $  0.106$ & $  0.108$ \\
  $\pi_{\e,\parallel}$  &...&...& $ -0.047$ & $ -0.034$ & $ -0.030$ & $ -0.042$ \\
                        &...&...& $  0.085$ & $  0.088$ & $  0.092$ & $  0.091$ \\
  $\pi_{\e,\perp}$      &...&...& $  0.344$ & $ -0.141$ & $  0.212$ & $ -0.573$ \\
                        &...&...& $  0.859$ & $  0.959$ & $  0.923$ & $  0.817$ \\
  $\psi$ (deg)          &...&...& $   92.5$ & $   93.1$ & $   93.3$ & $   94.1$ \\
    $I_{\rm S, KMTC01}$ & 21.01 & 21.00 & 21.09 & 21.07 & 21.07 & 21.07 \\
                    &   0.07 & 0.07 &  0.10 &  0.10 &  0.10 &  0.10 \\

    \hline
    \hline
    \end{tabular}
    \tablecomments{
The upper limits on $\rho$ are at $3\sigma$.}
    \label{tab:1492parms}
\end{table}

\begin{table}[htb]
    \renewcommand\arraystretch{1.05}
    \centering
    \caption{2L1S Parameters for OGLE-2018-BLG-0977}
    \begin{tabular}{c|r|r}
    \hline
    \hline
    Parameter &  Value & Error\\
    \hline
    $\chi^2$/dof  & $1553.0/1553$ \\
    \hline
    $t_{0}$  & 8276.600 
            &     0.018 \\
    $u_{0}$  & 0.1470 
            & 0.0097 \\
    $t_\e$ (days) & 20.37 
                  & 1.15 \\
    $s$  & 0.8793  
         & 0.0066 \\
    $q$ ($10^{-5}$) & 4.15 
                   & 0.43 \\
    $\alpha$ (rad) & 2.5063 
                   & 0.0079 \\
    $\rho$ & $<$0.0033 \\
    $I_{\rm S, KMTC01}$ & 20.268 
                    &  0.077 \\
    \hline
    \hline
    \end{tabular}
    \tablecomments{
The upper limit on $\rho$ is at $3\sigma$.}
    \label{tab:0977parms}
\end{table}

\begin{table}[htb]
    \renewcommand\arraystretch{1.05}
    \centering
    \caption{2L1S Parameters for KMT-2019-BLG-0953}
    \begin{tabular}{c|r|r}
    \hline
    \hline
    Parameter &  Value & Error\\
    \hline
    $\chi^2/dof$  & $909.0/9092$ \\
    \hline
    $t_{0}$  & 8638.149 
               & 0.015 \\
    $u_{0}$  & 0.491 
             & 0.023 \\
    $t_\e$ (days) & 11.62 
                  & 0.39 \\
    $s$  & 0.736 
         & 0.010 \\
    $q$ ($10^{-5}$) & 17.8 
                    & 3.4 \\
    $\alpha$ (rad) & 2.3601 
                   & 0.0077 \\
    $\rho$ & $<$0.028 \\
    $I_{\rm S, KMTC02}$ & 18.474 
                    & 0.078 \\
    \hline
    \hline
    \end{tabular}
    \tablecomments{The upper limits on $\rho$ are at $3\sigma$.}
    \label{tab:0953parms}
\end{table}

\begin{deluxetable}{lrrrrrr}
\tablecolumns{6} \tablewidth{0pc}
\tablecaption{\textsc{CMD Parameters for Six $q<2\times 10^{-4}$ Events}}
\tablehead{\colhead{Parameter} &
\colhead{KB190253} &
\colhead{OB180506} &
\colhead{OB180516} &
\colhead{OB191492} &
\colhead{OB180977} &
\colhead{KB190953}}
\startdata
$(V-I)_{\rm s}$ & 2.92$\pm$0.01 & 4.59$\pm$0.07 & 2.07$\pm$0.02& 1.48$\pm$0.04 & 2.17$\pm$0.05 & 2.05$\pm$0.04\\
$(V-I)_{\rm cl}$ & 3.31$\pm$0.02 & 4.47$\pm$0.02 & 2.58$\pm$0.02 & 1.84$\pm$0.02& 2.45$\pm$0.03& 2.56$\pm$0.03 \\
$(V-I)_{\rm cl,0}$ & 1.06 & 1.06 & 1.06 & 1.06 & 1.06 & 1.06\\
$(V-I)_{\rm s,0}$ & 0.67$\pm0.03$ & 1.18$\pm$15.10 & 0.55$\pm$0.03 & 0.70$\pm$0.05 & 0.78$\pm$0.05& 0.55$\pm$0.05 \\
$I_{\rm s}$  & 19.80$\pm$0.02 & 19.37$\pm$0.02 & 19.45$\pm$0.02 & 21.10$\pm$0.10 & 20.20$\pm$0.08& 18.46$\pm$0.08 \\
$I_{\rm cl}$ & 16.95$\pm$0.03 & 18.82$\pm$0.04 & 16.35$\pm0$0.03 & 15.40$\pm$0.04 & 16.12$\pm$0.03 &16.16$\pm$0.03 \\
$I_{\rm cl,0}$ & 14.44 & 14.55 & 14.48 & 14.38 & 14.47 & 14.38\\
$I_{\rm s,0}$ & 17.29$\pm$0.03 & 15.10$\pm$0.05 & 17.58$\pm$0.04 & 20.08$\pm$0.11 & 18.55$\pm$0.10 &16.68$\pm$0.09\\
$\theta_*$ ($\muas$) & 1.05$\pm$0.07 & 5.11$\pm$0.35 & 0.81$\pm$0.05 & 0.30$\pm$0.04 & 0.66$\pm$0.08&1.22$\pm$0.08 \\
$\theta_\e$ (mas) & $>0.23$ & $>0.43$ & $>0.08$& $>0.03$ & $>0.20$ & $>0.044$\\
$\mu_\rel$ ($\masyr$) & $>1.5$ & $>6.5$ & $>1.1$& $>0.2$ & $>3.6$ & $>1.4$ \\
\enddata
\tablecomments{Event names are abbreviations for, e.g.,
OGLE-2018-BLG-0506 and KMT-2019-BLG-0253. Lower limits on
$\theta_\e$ and $\mu_\rel$ are $3\,\sigma$.}
\label{tab:cmd}
\end{deluxetable}

\begin{table}[htb]
\renewcommand\arraystretch{1.5}
\centering
\caption{Bayesian Estimates of Physical parameters}
\begin{tabular}{c|c c c c c}
\hline
\hline
Event Name & $M_{\rm host}[M_{\odot}]$ & $M_{\rm planet}[M_{\oplus}]$ & $D_{\rm L}$[kpc] & $a_{\perp}$[au] & $\mu_{\rm rel}$[mas] \\
\hline
KMT-2019-BLG-0253 & $0.70_{-0.31}^{+0.34}$ &  $9.2_{-4.1}^{+5.0}$ & $4.9_{-1.6}^{+1.9}$ & $3.1_{-0.9}^{+0.8}$ & $4.7_{-2.1}^{+2.8}$ \\

OGLE-2018-BLG-0506 & $0.63_{-0.32}^{+0.37}$ &  $16.3_{-8.5}^{+12.0}$ & $5.6_{-1.6}^{+1.4}$ & $3.0_{-0.8}^{+0.7}$ & $8.1_{-1.5}^{+2.5}$ \\

OGLE-2018-BLG-0516 & $0.47_{-0.25}^{+0.38}$ &  $20.1_{-10.7}^{+16.4}$ & $7.0_{-2.2}^{+0.9}$ & $2.2_{-0.7}^{+0.8}$ & $5.0_{-1.9}^{+2.9}$ \\

OGLE-2019-BLG-1492 & $0.68_{-0.34}^{+0.34}$ &  $37.1_{-19.5}^{+25.1}$ & $5.6_{-2.2}^{+1.7}$ & $2.7_{-0.9}^{+0.9}$ & $3.8_{-1.7}^{+3.2}$ \\

OGLE-2018-BLG-0977 & $0.47_{-0.27}^{+0.38}$ &  $6.4_{-3.7}^{+5.2}$ & $6.5_{-2.2}^{+1.2}$ & $2.0_{-0.5}^{+0.6}$ & $6.9_{-1.8}^{+3.0}$  \\
KMT-2019-BLG-0953  & $0.28_{-0.17}^{+0.32}$ & $15.5_{-9.7}^{+19.0}$& $7.2_{-1.4}^{+0.7}$ & $1.1_{-0.4}^{+0.5}$ & $7.0_{-2.5}^{+3.3}$ \\
\\
\hline
\hline
\end{tabular}\\
\label{tab:phys}
\end{table}


\begin{deluxetable}{llccl}
\tablecolumns{5} \tablewidth{0pc}
\tablecaption{\textsc{Recovered Planets in KMT Prime Fields for 2018-2019}}
\tablehead{\colhead{Event Name} &
\colhead{KMT Name} &
\colhead{$\log q$} &
\colhead{$s$} &
\colhead{Reference} }
\startdata
OB181185 & KB181024 & $-4.17$ & 0.96 & \citet{ob181185} \\
KB181025$^a$ & KB181025 & $-4.03$ & 0.95 & \citet{kb181025} \\
OB180532 & KB181161 & $-4.01$ & 1.01 & \citet{ob180532} \\
OB180596 & KB180945 & $-3.74$ & 0.51 & \citet{ob180596} \\
OB181269 & KB182418 & $-3.24$ & 1.12 & \citet{ob181269} \\
OB180932 & KB182087 & $-3.15$ & 0.74 & in prep \\
OB180567 & KB180890 & $-2.91$ & 1.81 & \citet{ob180567} \\
KB180748 & KB180748 & $-2.69$ & 0.94 & \citet{kb180748} \\
OB180962 & KB182071 & $-2.62$ & 1.25 & \citet{ob180567} \\
OB180100$^a$ & KB182296 & $-2.58$ & 1.30 & in prep \\
OB181011$^{b,c}$ & KB182122 & $-2.02$ & 0.75 & \citet{ob181011} \\
OB181700$^b$ & KB182330 & $-2.00$ & 1.01 & \citet{ob181700} \\
OB181647$^b$ & KB182060 & $-1.95$ & 1.27 & in prep \\
OB181011$^c$ & KB182122 & $-1.82$ & 0.58 & \citet{ob181011} \\
OB181544 & KB180787 & $-1.60$ & 0.50 & Han et al.\ in prep \\ 
\hline
\hline
KB190842 & KB190842 & $-4.39$ & 0.98 & \citet{kb190842} \\
KB191953$^d$ & KB191953 & $-2.72$ & 2.30 & \citet{kb191953} \\
OB191180 & KB191912 & $-2.28$ & 1.90 & Chung et al.\ in prep \\
KB191552 & KB191552 & $-2.17$ & 0.74 &  in prep\\
OB190954 & KB193289 & $-1.77$ & 0.74 & \citet{ob190954} \\
OB190825 & KB191389 & $-1.63$ & 3.78 & Sato et al.\ in prep \\
KB193301 & KB193301 & $-1.40$ & 1.47 & Han et al.\ in prep \\
KB190371$^e$ & KB190371 & $-1.10$ & 0.83 & \citet{kb190371} \\
\enddata
\tablecomments{Event names are abbreviations for, e.g.,
OGLE-2018-BLG-1185 and KMT-2018-BLG-1024.  All $(s,q)$ from
``in prep'' events should be regarded as preliminary.
a: large $q$ degeneracy at $\Delta\chi^2<10$, or very large $q$ error.
b: $s$ degeneracy. 
c: Two-planet system.
d: identified as FSPL, but not investigated to find planet.
e: inner/outer degeneracy; with factor 1.5 higher $q$.}
\label{tab:2018recover}
\end{deluxetable}

\begin{deluxetable}{llrrl}
\tablecolumns{5} \tablewidth{0pc}
\tablecaption{\textsc{$\Delta\chi^2_0$
for $q<2\times 10^{-4}$ Sample Planets}}
\tablehead{\colhead{Event Name} &
\colhead{KMT Name} &
\colhead{$\Delta\chi^2_0$} &
\colhead{$\log q$} &
\colhead{Method} }
\startdata
OB191492 & KB193004 &   87 & $-3.75$ & Discovery  \\
OB180977 & KB180728 &  117 & $-4.38$ & Discovery  \\
OB180506 & KB180835 &  163 & $-4.09$ & Discovery  \\
OB180516 & KB180808 &  197 & $-3.93$ & Discovery  \\
KB190253 & KB190253 &  260 & $-4.39$ & Discovery  \\
KB190953 & KB190953 &  283 & $-3.75$ & Discovery  \\
OB191053 & KB191504 &  302 & $-4.90$ & Discovery  \\
KB190842 & KB190842 &  585 & $-4.39$ & Recovery   \\
OB181185 & KB181024 &  945 & $-4.17$ & Recovery   \\
OB180596 & KB180945 & 1828 & $-3.74$ & Recovery   \\
OB180532 & KB181161 & 7657 & $-4.01$ & Recovery   \\
\enddata
\tablecomments{Event names are abbreviations for, e.g.,
OGLE-2018-BLG-1185 and KMT-2018-BLG-1024.}
\label{tab:dchi20}
\end{deluxetable}

\clearpage

\epsscale{0.9}

\begin{figure}
\plotone{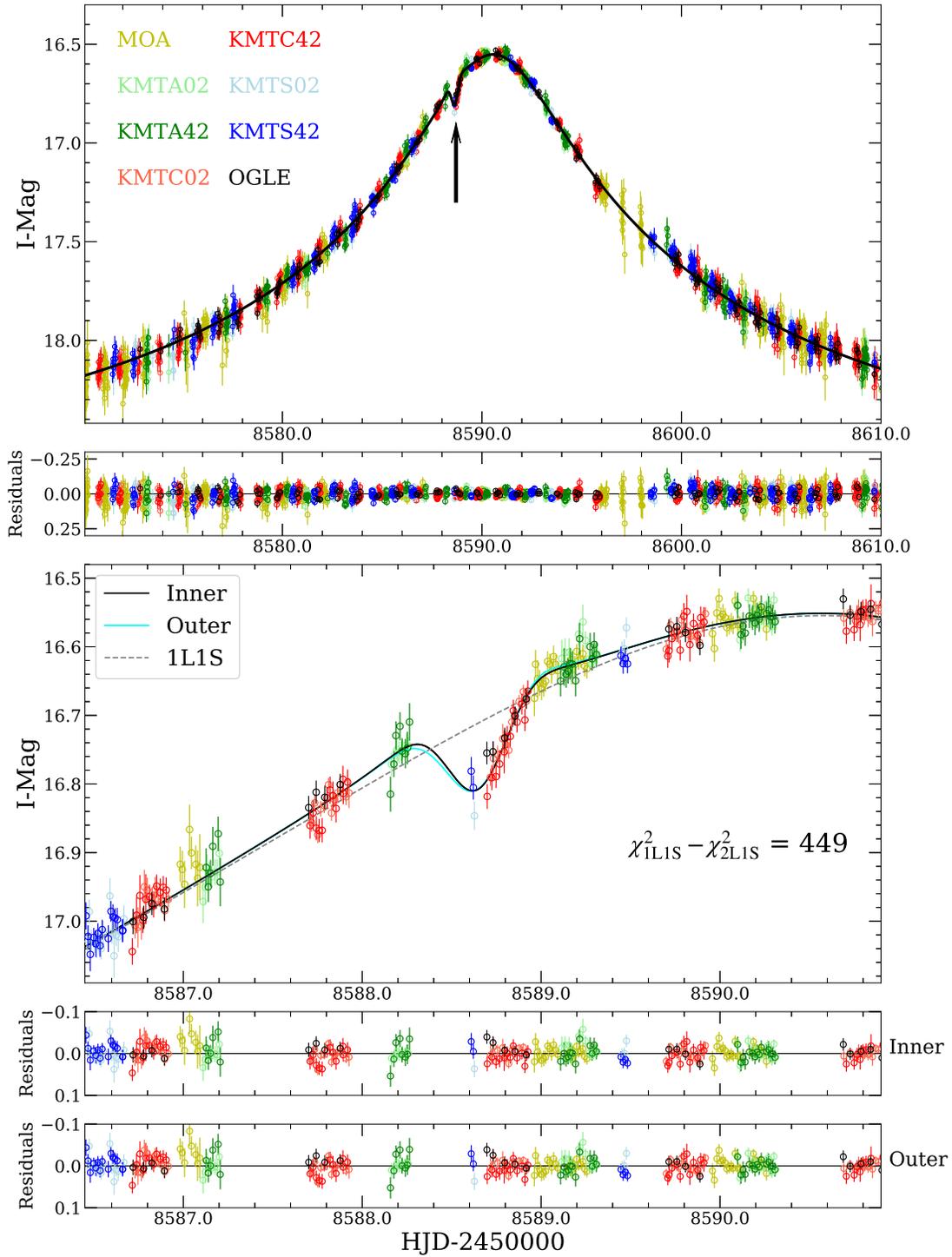}
\caption{Light curve and models of KMT-2019-BLG-0253.
The main evidence for the ``inner'' and ``outer'' planetary model
(which are virutally identical) is the ``dip'' in KMTC data from Chile,
near 8588.6.   This dip is supported by contemporaneous
OGLE data, also from Chile, and also by two points from KMTS in South Africa.
The dip is the signature of a minor-image perturbation,
passing either ``inside'' or ``outside'' the two triangular
caustics associated with the minor image.  See Figure~\ref{fig:4caustics}.
}
\label{fig:0253lc}
\end{figure}

\begin{figure}
\plotone{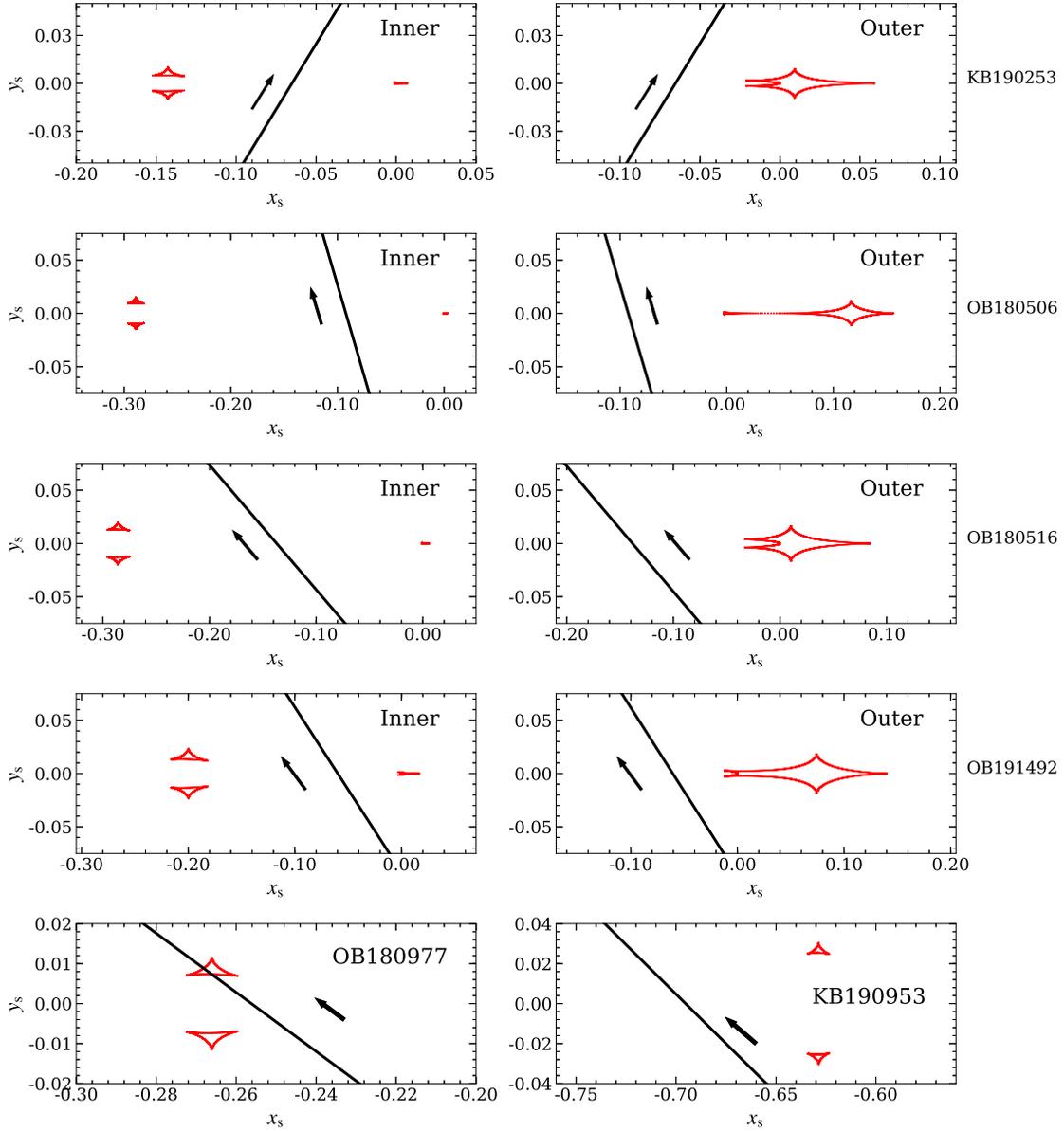}
\caption{Caustic topologies for all {6} events.
For the first 4,
the degenerate pairs have identical topologies.  In each case, for the ``inner''
topology, the source passes between the central caustic and the
two triangular caustics associated with the minor image (opposite 
side of host from the planet).  The ``dip'' is due to a magnificatoin
trough that runs along the planet-host axis, between these caustics.
For the ``outer'' topology, these caustics have merged with the
central caustic to form a resonant caustic.  The ``dip'' is then due
to the magnification trough that extends out from between these two wings
of the resulting resonant caustic.   For OGLE-2018-BLG-0977,
the source intersects the triangular caustics, so there is no degeneracy.
{For KMT-2019-BLG-0953, there is only an ``outer'' solution.
See Section~\ref{sec:static-ob190953}}.
}
\label{fig:4caustics}
\end{figure}

\begin{figure}
\plotone{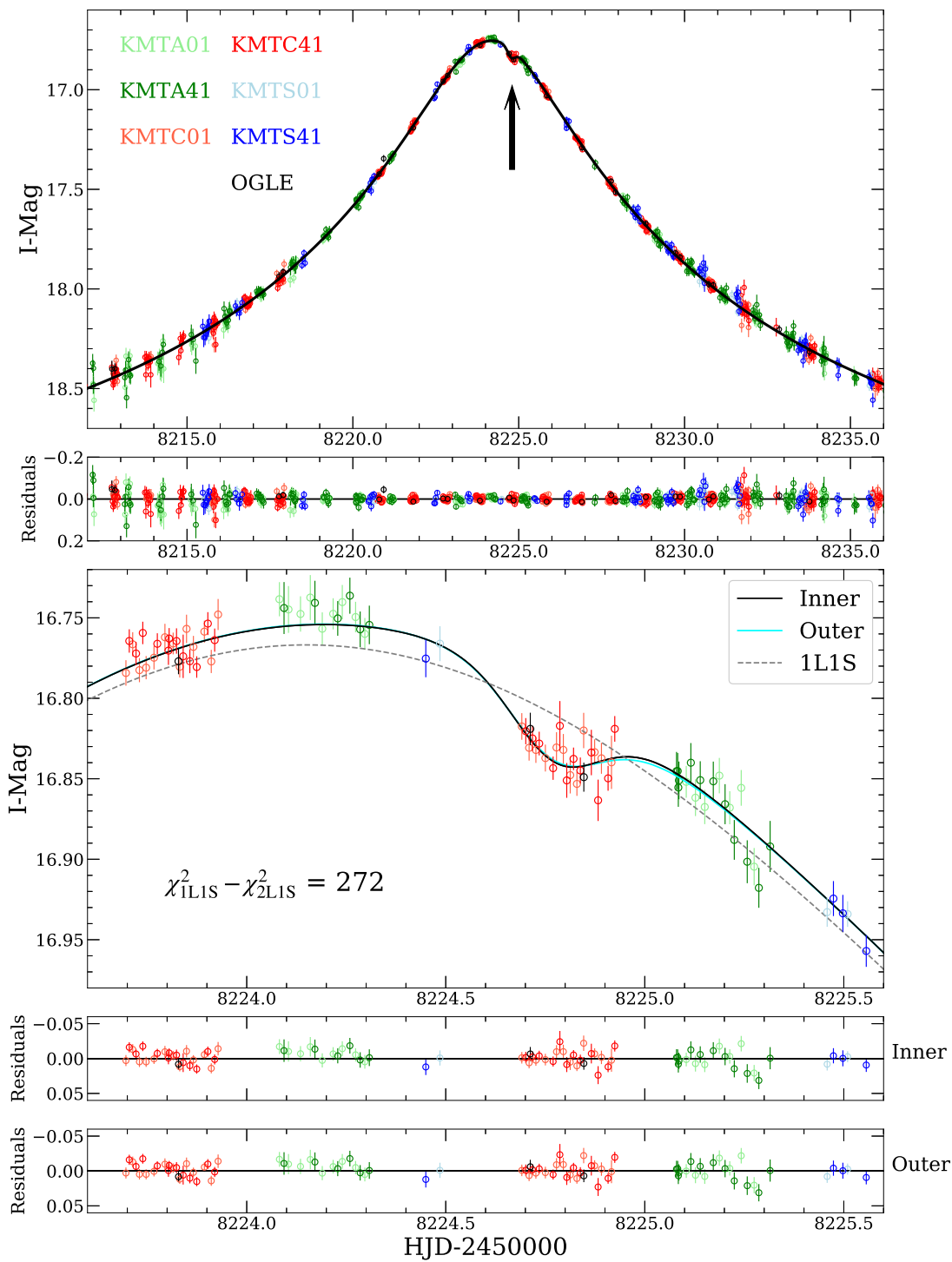}
\caption{Light curve and models of OGLE-2018-BLG-0506.
The main evidence for an anomaly comes from the dip at about 8224.7
as seen in 26 KMTC points.  This is confirmed by two OGLE points.
The caustic topology is the same
as for KMT-2019-BLG-0253.  See Figure~\ref{fig:4caustics}.
}
\label{fig:0506lc}
\end{figure}

\begin{figure}
\plotone{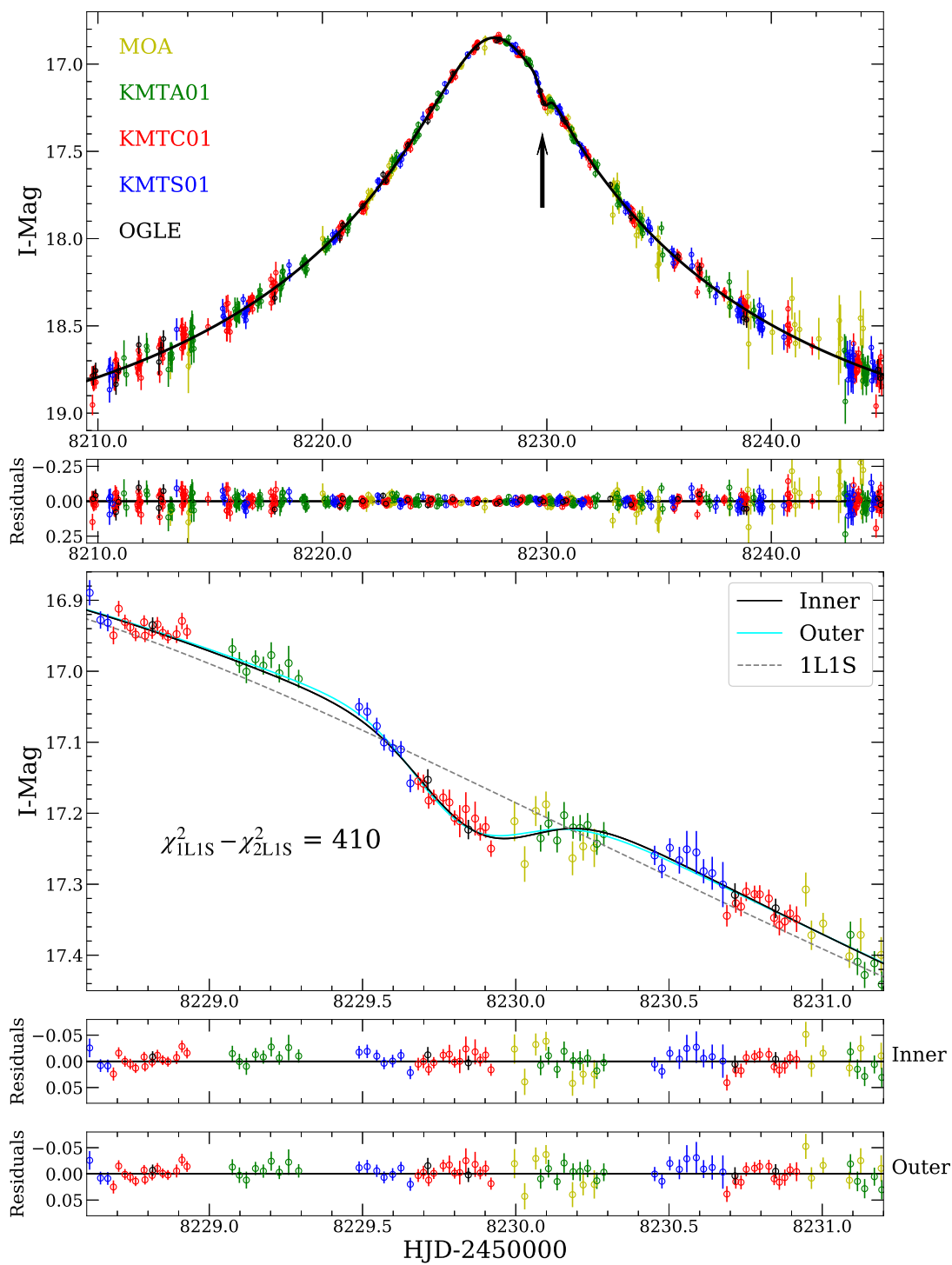}
\caption{Light curve and models of OGLE-2018-BLG-0516.
The main evidence for an anomaly comes from the dip at about 8229.85 .
This is confirmed by two OGLE points.
The KMTA and MOA
data trace the end of the dip, while the KMTS data are consistent
with the beginning of the dip.  The caustic topology is the same
as for KMT-2019-BLG-0253 and OGLE-2018-BLG-0506.  
See Figure~\ref{fig:4caustics}.
}
\label{fig:0516lc}
\end{figure}

\begin{figure}
\plotone{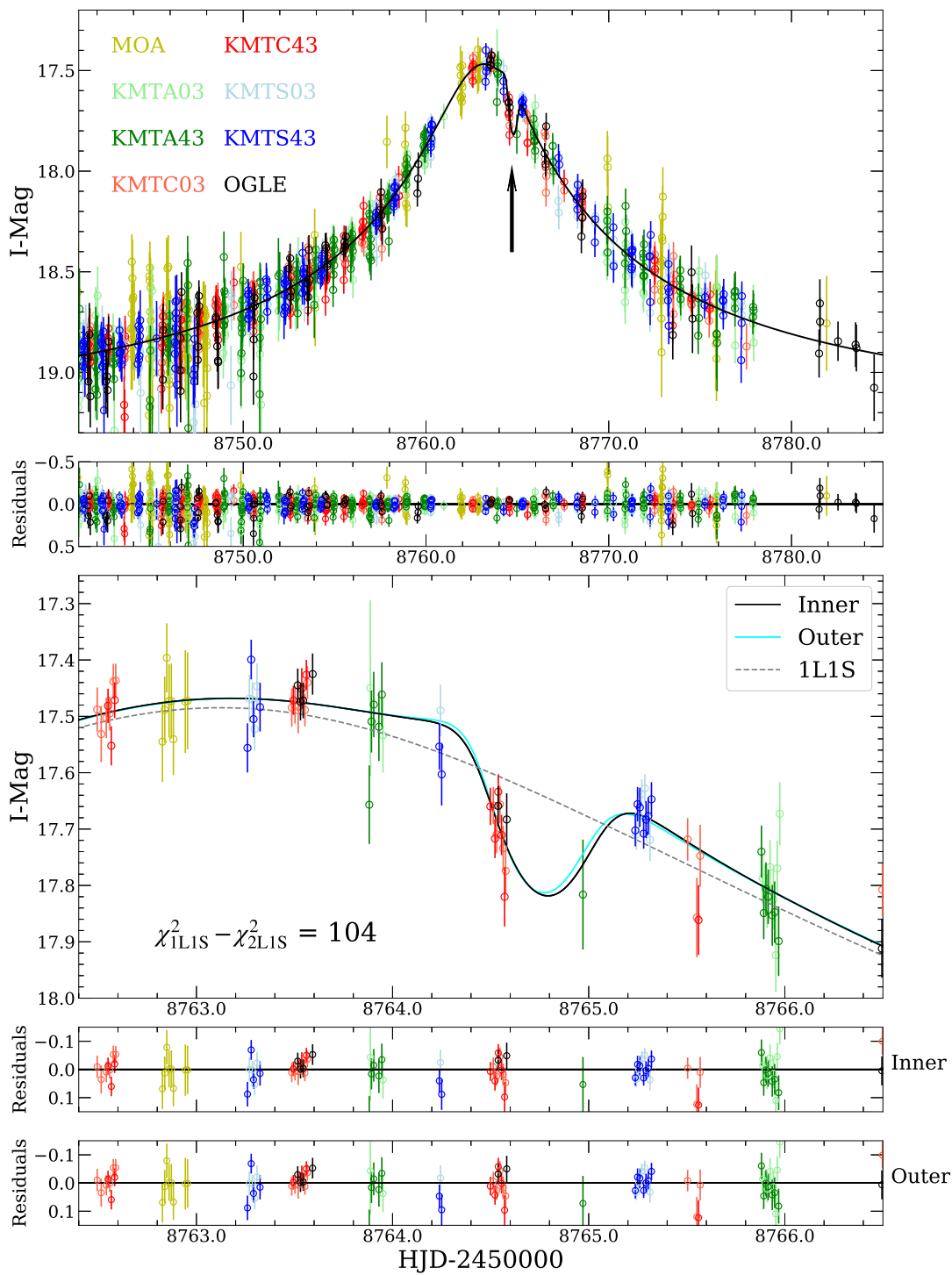}
\caption{Light curve and models of OGLE-2019-BLG-1492.
The main evidence for the anomaly comes from the dip in KMTC
data at about 8764.65, which is confirmed by two OGLE points. 
It is also supported by the post-dip ``ridge'' in
KMTS data at about 8665.25.  The caustic topology is the same
as for KMT-2019-BLG-0253, OGLE-2018-BLG-0506, and OGLE-2018-BLG-0516.
See Figure~\ref{fig:4caustics}.
}
\label{fig:1492lc}
\end{figure}

\begin{figure}
\plotone{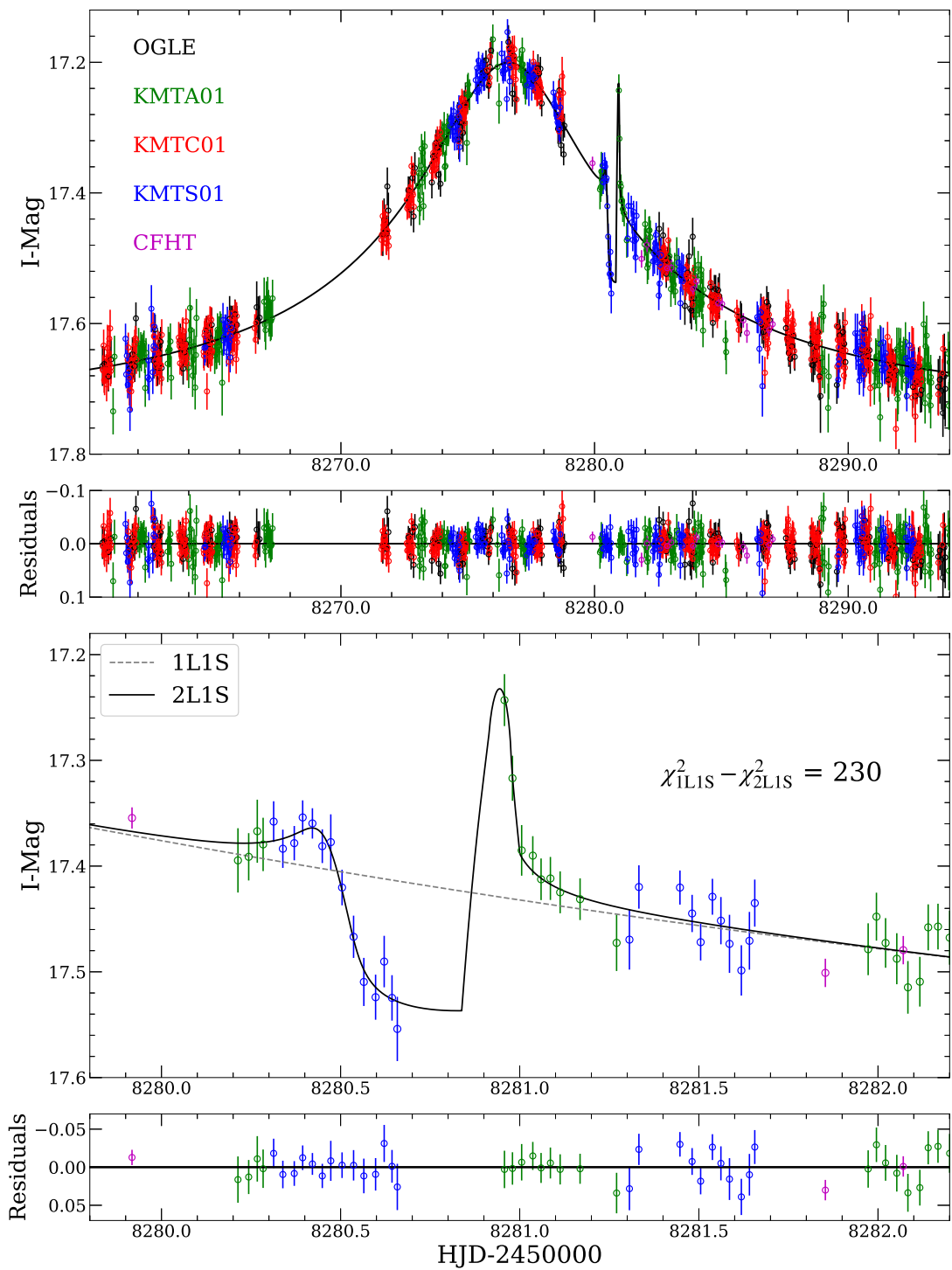}
\caption{Light curve and model of OGLE-2018-BLG-0977.
The KMTS data near 8280.5 briefly rise and then sharply
fall, which indicates entrance into the trough that threads
the two minor-image triangular caustics, near (but not over)
one of these caustics.  The KMTA data near 8281.0 show a sharp
decline, followed by flattening, consistent with an exit
from the triangular caustic on the opposite side of the trough.
The caustic-topology diagram (Figure~\ref{fig:4caustics}) confirms this 
qualitative light-curve analysis.
}
\label{fig:0977lc}
\end{figure}

\begin{figure}
\plotone{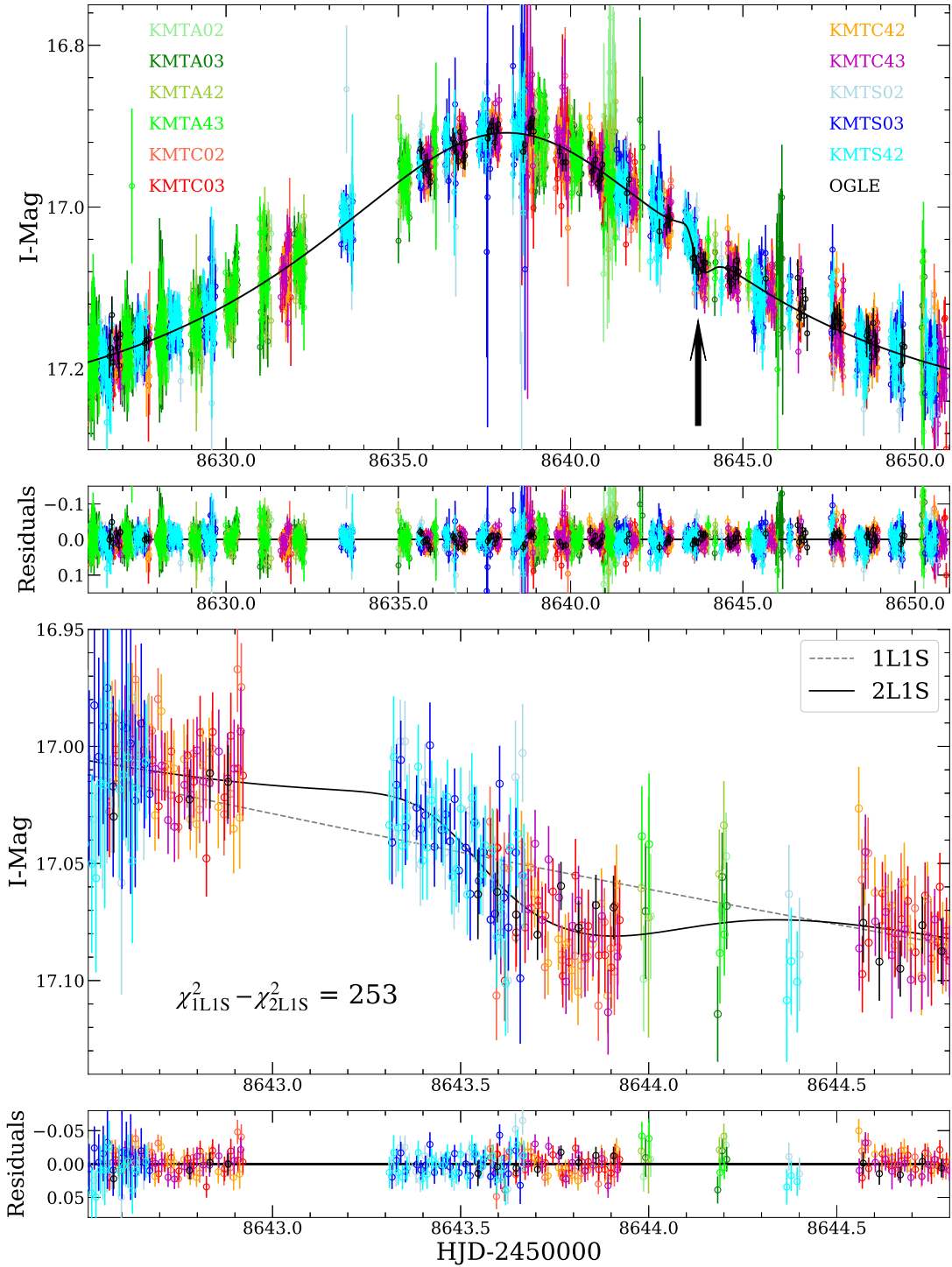}
\caption{Light curve and model of KMT-2019-BLG-0953. The KMTS and 
KMTC data show a low-amplitude dip $8643.4 \la {\rm HJD}^\prime < 8643.9$,
which is not much 
larger than the individual error bars, but because the event lies 
in a small region with very high cadence, there are $>100$ data 
points during this 12 hour interval.  Eight OGLE points near the 
trough confirm the dip. The dip indicates that the source passed 
near, but not over the minor image due to the host. See
Figure~\ref{fig:4caustics}.}
\label{fig:0953lc}
\end{figure}

\begin{figure}
\plotone{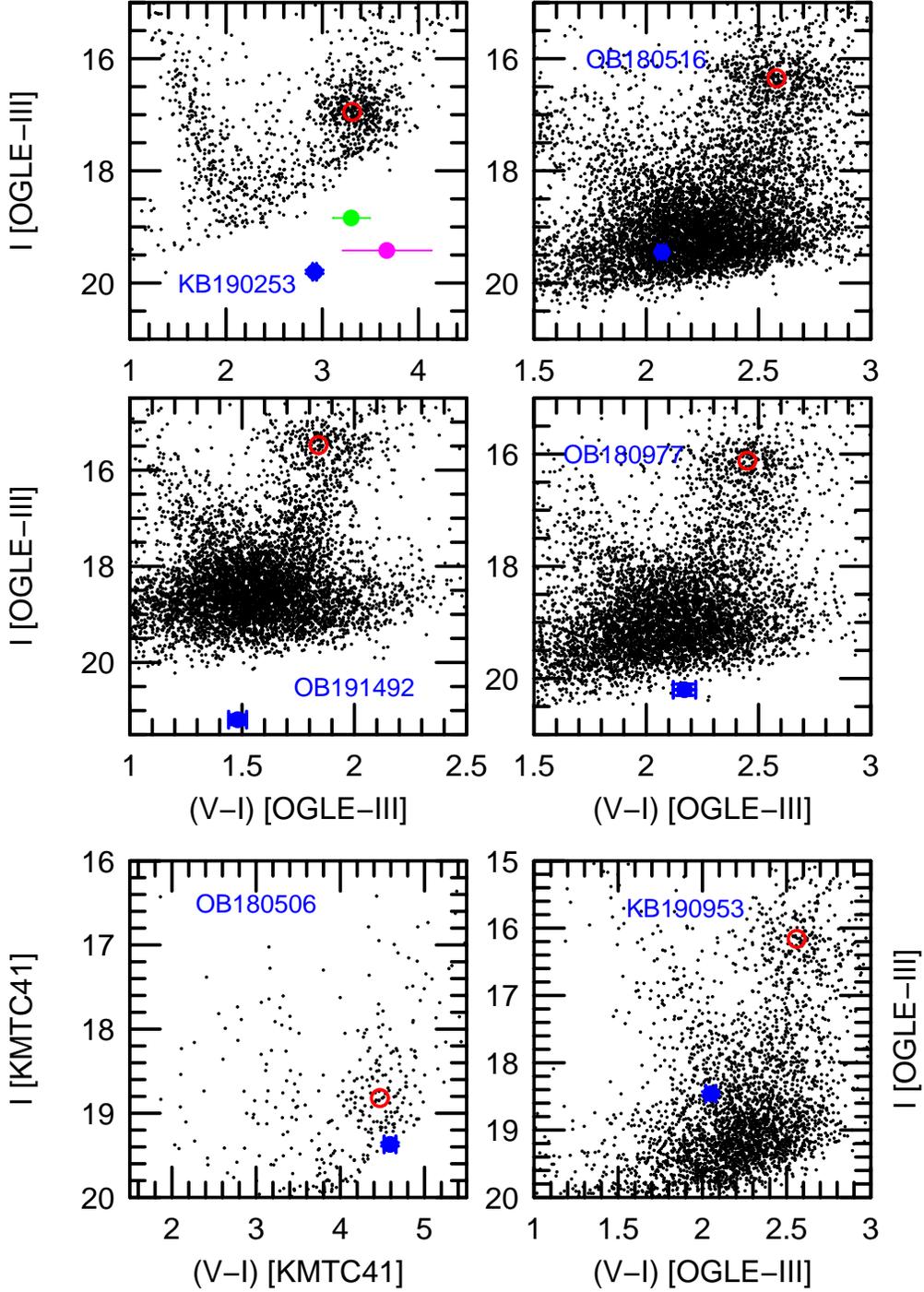}
\caption{CMDs for each of the {six} $q<2\times 10^{-4}$ planets reported here.
The source positions (blue) and clump-giant centroids (red) are shown
for all events.  For KMT-2019-BLG-0253, the
baseline object (green) and blended light (magenta) are also shown.
Photometry is in the calibrated OGLE-III system for {all except
OGLE-2018-BLG-0506 (lower left), which is in}
the KMTC41 pyDIA system.
}
\label{fig:allcmd}
\end{figure}

\begin{figure}
\plotone{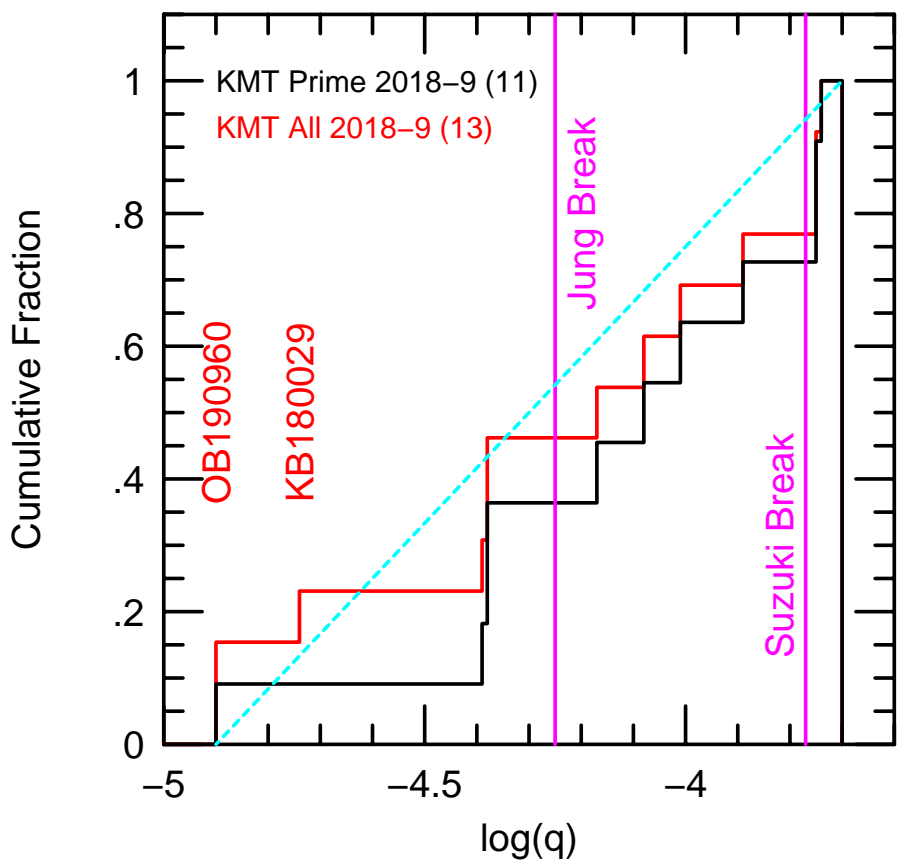}
\caption{Cumulative distribution of recovered and discovered KMT
prime-field planets from 2018-2019 (black) compared to 
the mass-ratio-function breaks proposed by \citet{suzuki16}
and \citet{kb170165} (magenta).  The cyan dashed line 
simply connects the first and last KMT detections.
If the underlying distribution of mass ratios
(after multiplying by the detection efficiency) were uniform in
$\log q$, the cumulative distribution would approximately
follow this line.  The red event names at left indicate the mass ratios 
of the only known $q<2\times 10^{-4}$ planets from 2018-2019 non-prime fields.
If no other such planets are found when the AnomalyFinder is applied
to these fields, the cumulative distribution will be as shown in red.
}
\label{fig:cum}
\end{figure}

\end{document}